\PassOptionsToPackage{numbers,square,sort&compress}{natbib}
\documentclass[10pt]{article}
\usepackage[a4paper,margin=24mm]{geometry}
\usepackage{amsmath,booktabs,tabularx,array,microtype,graphicx,xcolor,natbib,xurl,hyperref,placeins}
\setcitestyle{numbers,square,comma}
\usepackage[T1]{fontenc}
\usepackage[utf8]{inputenc}
\usepackage{pdfpages}
\usepackage{lmodern,caption}
\hypersetup{colorlinks=true,linkcolor=black,citecolor=blue!45!black,urlcolor=blue!45!black,pdfauthor={}}
\newcommand{\mainfigure}[3]{%
  \begin{figure}[!htbp]
    \centering\includegraphics[width=#2\linewidth]{#1}
    \caption{#3}
  \end{figure}}

\title{Who Moves the Deity? Mechanics, Participation, and Attributed Agency in Himalayan Rath Carrying}
\author{Nalin Dhiman$^{1,*}$\\[3pt]
\small $^1$School of Computing and Electrical Engineering,\\
\small Indian Institute of Technology Mandi, Kamand, Mandi,\\
\small Himachal Pradesh, India\\[2pt]
\small $^*$Correspondence: \texttt{d24008@students.iitmandi.ac.in}}
\date{}
\newcommand{\captionone}{A Kullu chair-type rath and its carrying arrangement. The original schematic is informed by Ha\d{d}imb\={a} Dev\={i} and Manu \d{R}\d{s}i's apparatus in Halperin (2016, pp.~14-15) and Luchesi (2006, pp.~70-72). Two longitudinal poles flank the carriers' heads and rest at shoulder level; a wooden frame supports cloth and schematic metal faces. Proportions, posture, decoration and hidden contacts are illustrative. Hands indicate handling, not calculated forces. Additional views appear in Supplementary Figure~S3.}
\newcommand{\captiontwo}{An assumed carrying arrangement and its mechanical reduction. (a) Four candidate shoulder contacts grouped into two vertical resultants, one per pole side. (b) Separate synthetic body with assumed centre of mass (COM), half-spacing $a=0.50$~m and COM height $h=0.15$~m above the contact line. These are not estimates from Figure~1. (c) Compression $\delta_i=b_i-z_i$ is a coordinate difference, not a physical pad thickness. Summed forces and moments yield the spring-damper model only under additional kinematic and constitutive conditions, derived in Supplementary Section~S2.}
\newcommand{\captionthree}{Support timing and the resulting motion. Columns compare (a) zero and (b) 60-degree offsets between equal 4-mm, 1.5-Hz inputs. Rows show prescribed support heights, solved heave and solved roll, with identical axes across each row. Traces cover 4-8~s; RMS values use 4-12~s inclusive, with heave measured about its sample mean and roll about zero. The 60-degree offset concerns input timing (0.111~s), not object tilt.}
\newcommand{\captionfour}{Walking-informed inputs and consequences in the synthetic four-contact extension. (a) Mean shoulder heights of front and rear carriers in G90: 90 degrees of stride phase gives opposed vertical motion. (b) Solved pitch and roll in G90; neither angle is prescribed. (c) Opposing individual left-side loads in G180, despite negligible roll and pitch; the dashed curve divides the reduced model's left resultant equally and does not recover the separate loads. (d) Summed left-side force in the larger-input stress case, comparing separate contact laws with aggregation before gating. Traces show 8-10~s; summary statistics use 8-24~s. These are assumed walking waveforms, not procession recordings.}
\begin{document}
\maketitle
\begin{abstract}
People carry ritual palanquins, or raths, while describing themselves as following a deity's movement. This study combines western Himalayan ethnography, motor-control research and synthetic mechanics to explain how practical contribution and attributed agency relate. It distinguishes knowing a destination, preparing for a load change, influencing local movement and attributing an episode's purpose. An oracle-mediated travel proposal connects divine initiative to practical responsibility; accounts of rehearsal, shoulder adjustment, upright carrying and a narrated return identify distinct forms of participation. Using assumed parameters and prescribed shoulder motions, a two-support model separates heave from roll, and a four-contact walking extension adds pitch. The extension demonstrates substantial changes in individual loads with negligible rocking and shows how pitch and contact opening affect load sharing. A worked comparison with an upright-carrying account explains how visible posture can conceal changing support. These findings connect ritual authority, bodily preparation and local physical influence while specifying evidence that distinguishes guidance, anticipation, correction and retrospective interpretation. The resulting account treats following as an organised form of participation: people can sustain and adjust a movement while attributing its initiative and purpose to the deity.
\end{abstract}

\section{Introduction}

In a story recorded by Halperin, devotees carrying the goddess Ha\d{d}imb\={a}'s rath to the Dasahra festival were running late. They passed a snack seller who offered tea and halwa, a sweet dish, as pras\={a}d (sacred food). Further along the journey, the rath stopped and returned to him. Rishi, the Old Manali resident recounting the episode, understood the return as the goddess teaching the devotees a lesson for ignoring the offer \citep[pp.~322-323]{halperin2026}. Rishi's retrospective account gives the interruption a purpose: the return corrects the devotees' conduct.

The account brings the paper's question into view. A rath is carried through the work of people who lift, balance and support it, yet those people can describe themselves as following its movement. What kinds of knowledge and control are involved? A person can know a destination without anticipating each pressure change, anticipate a load change without deciding the destination, or influence both while describing the event as the deity's action. Treating these as the same claim makes practical skill and reported following appear more contradictory than they need be. The question is especially consequential when a movement is understood to correct the participants' own intentions, as in the return narrative.

The object being carried is itself part of this relation. In the Kullu chair-type arrangement discussed here, poles, a dressed wooden frame and metal faces form an assembled manifestation of the deity. Halperin explains that a temple idol is not ordinarily placed inside this vehicle: the assembled rath fulfils the function of a divine image \citep[pp.~14-15]{halperin2016}. Luchesi situates Ha\d{d}imb\={a}'s appearances in assembly, decoration, attendance and encounters with worshippers \citep[pp.~70-72]{luchesi2006}. Participants accompany a socially and ritually constituted presence, as well as support a material load.

Existing scholarship already explains this conjunction. Halperin connects material construction and differentiated participation to complex agency, and uses Meyer's account of sensational forms to explain cultivated, tangible divine presence \citep{halperin2016,halperin2026,meyer2006}. Sutherland analyses palanquin movement within regional sovereignty and theistic subjection; Berti examines communication, authority and disagreement \citep{sutherland2006,berti2006}. Their explanations concern how divine action becomes authoritative and experienced within a practice. These analyses locate the work of carrying within established relations of presence, authority and obligation.

The central proposition of this article is that following can be an organised form of participation. An authorised initiative can establish what is to be done, while bearers prepare and adjust the support through which it is accomplished. Knowledge of a destination, preparation for a load, influence over a moment of motion and attribution of an episode's purpose therefore need to be distinguished. None, taken alone, establishes the others. This proposition leaves open both deliberate guidance and compelling experiences of being led.

The contribution has two connected parts. A comparison of published accounts specifies how ritual offices, practical knowledge and bodily response can contribute at different scales. A mechanical counterexample then shows why even a detailed observation of the object's orientation may fail to reveal individual load sharing. Together they identify a problem of evidence: a public action, its local physical production and participants' accounts of authorship require related but different observations. The argument connects these observations through the organisation of a task, the distribution of support and the attribution of its purpose.

The argument proceeds from accounts of participation to the material demands of the actions they describe. An oracle-mediated proposal to travel identifies how initiative can organise preparation. An account of keeping a heavy rath upright identifies a controlled posture whose appearance can conceal the work of support. The numerical examples establish consequences of specified inputs for heave, roll, pitch and load sharing. Their role is to test an inference about visible movement and contribution within an account whose ritual specificity comes from the sources.

\section{Knowledge, preparation and local control}

\subsection{What it can mean to follow}

Four questions organise the comparison. What does a person expect to happen? What action do they prepare? Which aspect of motion does their action influence? What source or purpose do they assign to the episode? An intention to pass safely through a gate, an anticipated shift of weight, a shoulder adjustment and an account of the deity's initiative answer different questions. They may belong to the same person and occur within the same passage.

Physical influence and practical control are related but different. An adjustment has physical influence if changing it, with the relevant other conditions held fixed, would change the load's motion. Practical control concerns using adjustments to maintain or change something: footing, height, tilt or direction. A bearer may help keep the object upright without being able to choose its destination. Control therefore needs an object and a timescale. Maintaining balance over a moment and determining the course of a procession are different achievements, even when the same person participates in both.

Experienced authorship adds a further question: what does the person feel or judge themselves to have brought about? Joint-action research distinguishes experiences of one's own contribution, a partner's contribution and a shared outcome \citep{loehr2022}. The present argument uses this distinction to ask more precise questions about carrying. An account of the deity's initiative can also concern legitimate authority, appropriate conduct and the significance of an encounter. Those matters require the ethnographic interpretation that a force record cannot provide.

Consider a simple analytical example, not a reported field event. Two people intend to carry a load through an entrance. Both know where they are going; one adjusts a shoulder after feeling a change in pressure, and the other adjusts while expecting that change. Their shared destination does not identify which local adjustment came first. Nor would a recording of the resulting tilt alone distinguish their preparation. This example shows why an account of following need not imply the absence of every intention. It also leaves room for a person to experience compulsion or a loss of control: neither that experience nor its absence follows from the existence of a physical contribution.

\subsection{Learning to expect and to adjust}

Familiarity can contribute to each question differently. Halperin describes children learning to attend to an approaching rath and distinguishes customary stops from contingent encounters \citep[pp.~315-317]{halperin2026}. Someone may therefore recognise an arrival without knowing how to carry. Differentiated responsibilities in Luchesi's account likewise create different opportunities to assemble, handle, observe and interpret the apparatus \citep[pp.~70-72]{luchesi2006}. The distinction directs attention to the specific handling experience available through each role.

Practical carrying gives learned anticipation a specific possible role. Changing bearers, matching shoulder height, protecting the assembled structure and negotiating entrances require preparation for changing loads \citep[pp.~318-320]{halperin2026}. The hypothesis is that relevant experience can improve such preparation beyond what a present instruction or visible cue alone specifies. Knowing that a doorway is narrow may support an explicit plan. Preparing for a familiar pressure change concerns a more local demand. Successful prediction by itself does not show that the predictor caused the motion.

Differences between bearers could organise the team rather than average away. A person with relevant experience might anticipate an obstacle, issue a useful instruction or compensate for a partner's movement. Height, strength, fatigue and familiarity could affect different parts of this process. A possible guide should be identified through what they knew, communicated and changed, rather than inferred from a larger force or greater age alone. The model's unequal shoulder waveforms test consequences of input differences; they do not encode these personal capacities.

Laboratory manipulation makes the proposed distinction concrete. Johansson and Westling varied weights hidden from view during precision-grip lifting; forces before lift-off reflected the preceding weight, and unexpected changes revealed mismatched preparation \citep{johansson1988}. Flanagan and colleagues combined force-sensor lifts of objects with an inverted size-weight relationship with judgments of equally weighted test objects. Lifting forces adapted faster than weight judgments \citep{flanagan2008}. These studies separate action preparation from a perceptual report. Perceived weight is not experienced authorship: extending this distinction to rath participation is the present hypothesis, not their measured result.

Other people's movements can provide useful advance information. In Aglioti and colleagues' basketball task, elite players predicted outcomes earlier than novices and expert watchers \citep{aglioti2008}. The comparison involved existing groups, and some watchers had played previously. It therefore motivates attention to relevant experience and available cues without isolating training experimentally or identifying a carrier-specific mechanism. Rapid correction can also incorporate task goals and constraints \citep{pruszynski2011,yang2011}. Preparation, intention and sensory response are compatible components of skilled action.

Applied to carrying, this reasoning suggests that familiarity should be specified by what has been learned. Recognising a customary stop, anticipating a particular load and knowing a partner's handling style could each support different preparation. This distinction makes a useful empirical difference: a person's accurate expectation of the next stop would not, by itself, demonstrate apparatus-specific motor learning. Conversely, an effective local adjustment need not depend on knowing the episode's eventual purpose.

\subsection{Evidence that distinguishes the explanations}

\begin{table}[!htbp]
\centering\small
\caption{Candidate explanations and the evidence that would distinguish their roles. These are analytical alternatives that may coexist, not fitted or mutually exclusive models.}
\label{tab:explanations}
\begin{tabularx}{\linewidth}{@{}>{\raggedright\arraybackslash}p{.21\linewidth}>{\raggedright\arraybackslash}p{.30\linewidth}>{\raggedright\arraybackslash}X@{}}
\toprule
Explanation & Specific proposal & Discriminating evidence\\
\midrule
Deliberate guidance or common instruction & Participants intend an agreed action or respond to an instruction. & Recorded instructions, intended goals and who could perceive them, linked to subsequent action; agreement on a destination does not specify local force timing.\\
Learned anticipation & Relevant prior handling prepares an adjustment before the load change is fully encountered. & Preparation varying with handling history or apparatus familiarity when current cues and instructions are comparable. A temporal lead alone is insufficient.\\
Online correction & Adjustment responds to information available through the load or other people. & A response linked to an independently identified sensory change, with earlier cues also considered. Response does not exclude prior preparation.\\
Retrospective interpretation & An account assigns purpose or authorship to an episode. & Comparison of advance expectations, immediate accounts and later narratives; a later narrative need not identify the initiating action.\\
\bottomrule
\end{tabularx}
\end{table}

Table~\ref{tab:explanations} makes familiarity a discriminating proposal rather than a synonym for participation. Comparing handling with observation could identify different preparation, but differences in strength, role, partners and instructions would also need attention. Comparing familiar with unfamiliar apparatus while keeping cues comparable could test specificity; changed preparation would not establish diminished control. Comparing expectations with immediate and later accounts could distinguish prospective knowledge from retrospective purpose, but agreement across reports would not identify the force sequence. The comparison thereby specifies how observations of familiarity can discriminate among explanations.

\section{What the ethnographic contrasts establish}

Four kinds of material help locate these explanations within the sources. First, Sutherland describes boys making toy palanquins in Rohru and Tukpa. He asked one group to stage a performance for his camera, after which adults allocated roles and directed a simplified ceremony \citep[p.~100]{sutherland2006}. The episode demonstrates rehearsal of a public repertoire under explicit direction. It supplies neither adult load-handling competence nor evidence that a rehearsed movement became involuntary. Current instruction and opportunities for learning appear together, so the account cannot isolate either as the cause of a later action.

Second, Sutherland records an interlocutor explaining how a bearer lowers a shoulder when the palanquin feels heavier on that side, keeps footing and follows its movement \citep[p.~104]{sutherland2006}. This connects immediate sensation, practical adjustment and a bow recognised as respectful conduct. Sutherland places this account within theistic subjection and regional forms of sovereignty, rather than treating bodily response as an explanation that dissolves divine action. The account supports the intelligibility of responsive contribution. Its narrative ordering does not supply the timed force and sensory records needed to distinguish learned preparation from online correction.

Third, Halperin's discussion includes heavy raths kept upright because lateral swinging would endanger carriers \citep[pp.~320-322]{halperin2026}. Religious presence and a recognisable movement style therefore do not require conspicuous roll in every case. Keeping upright is a practical achievement to be explained, not an absence of meaningful action. This contrast directs attention to which movement is being controlled. Sutherland also relates regional styles to pole construction, but his longhair and Mah\={a}s\={u} forms are distinct from the Kullu apparatus illustrated here \citep[pp.~98-99]{sutherland2006}.

Fourth, the return narrative that opens this article concerns the purpose assigned to a whole episode \citep[pp.~322-323]{halperin2026}. Rishi's account makes the destination significant: returning to the overlooked seller corrects the devotees' conduct. The contrast with Sutherland's account concerns the action being explained: Rishi gives the journey's reversal a purpose, while Sutherland's interlocutor describes responding to a felt load.

Taken together, these materials separate rehearsal, felt load, controlled posture and narrated purpose. An account of following may concern immediate sensation, proper conduct, attributed initiative, or several at once. Reading every such account as one measure of reduced personal control would erase differences that the sources themselves make consequential. Conversely, familiarity with a movement style can enable recognition without implying that the observer generated it.

\subsection{Authoritative speech and practical responsibility}

The role of the g\={u}r, or deity's medium, requires attention to particular acts of speaking and receiving a message. In Halperin's account, \'{S}a\.{n}kh N\={a}r\={a}ya\d{n}'s wish to attend Kullu's Dasahra after a long absence is conveyed through his oracle. Halperin then records a conversation with Mahesh, a local priest, about where the visiting deity would stay. Mahesh assigns responsibility for seeking a place to the k\={a}rd\={a}r, the deity's manager, who would need to approach the district administration. He also anticipates uncertainty if the expected space is unavailable \citep[pp.~314-315]{halperin2026}.

The account separates an attributed initiative, its communication and the practical work expected of an office holder. The planning discussion makes the relationship between religious authorship and human planning concrete: the deity's wish provides a reason for action, while finding space becomes someone's responsibility. Interpretation is already involved in defining the task before the proposed journey takes place.

The offices mediating such relations vary. In Sutherland's Rohru and Tukpa material, the vaz\={i}r addresses questions to the deity in palanquin form; the oracle gives voice to the deity's orders, and the vaz\={i}r organises their execution by the community \citep[pp.~102-103]{sutherland2006}. The medium and the person issuing practical instructions therefore need not be the same office holder. Berti's Kullu account further places consultation within reciprocal obligations and contested claims to status \citep[pp.~45-48]{berti2006}. Authority involves relationships and possible disagreement, rather than automatic agreement among everyone present.

For the present argument, these distinctions identify how speech could affect movement: it could specify a destination, a greeting, a pause or another task to which bearers direct their preparation. Whether particular words changed a shoulder input would require evidence about who heard them, prior expectations and the action sequence. An instruction might initiate a response, confirm an anticipated action or offer an account of motion already underway. The sources thus locate communication within the distribution of practical responsibility.

\subsection{Material form and appropriate conduct}

Halperin's use of Meyer's concept of sensational forms helps explain why the apparatus matters beyond its weight. In that account, cultivated forms of worship organise bodily participation and access to divine presence \citep[pp.~306-307]{halperin2026}. Assembly, carrying and encounter give the object a place in an established practice. The poles enable a particular way of supporting the rath; the assembled faces and cloth help make the carried presence recognisable. Mechanical contact and religious significance belong to this shared setting, although they answer different explanatory questions.

The distinction has a consequence for the analysis of motion. A measured inclination is a change in orientation. Calling it a respectful bow additionally relates it to an encounter and a convention of conduct, as Sutherland's interlocutor does \citep[p.~104]{sutherland2006}. It would therefore be insufficient to match an angle trace and conclude that the model had reproduced a bow as a ritual action. Equally, identifying its significance would not specify the individual forces that produced the inclination. This is where the two forms of explanation can inform one another: the ethnography identifies the action that matters, while mechanics specifies some of its material requirements.

The resulting proposal is a relation that can run in both directions. Existing expectations can organise what people prepare to do, and an unexpected movement can become material for a later account. It is not a fixed sequence in which a meaningless oscillation occurs first and culture labels it afterward. The four explanations in Table~\ref{tab:explanations} concern distinguishable roles within an event, not four compulsory stages through which every procession passes.

\subsection{Ritual attention, affiliation and authorship}

Ritual theories clarify different parts of this relationship. Boyer and Li\'enard propose that prescribed performance can direct attention towards the details of an action and the importance of performing it correctly. Their distinction between ritualisation and routinisation allows repeated ceremonies to combine demanding, closely monitored acts with familiar routines \citep[pp.~610-611]{boyer2006}. Applied here, this suggests a hypothesis about what a bearer attends to: appropriate greeting or posture may require attention even when carrying is familiar. The distinction focuses attention on how familiarity and task demands shape attention during ritual performance.

Watson-Jones and Legare organise the social functions of group ritual around membership, commitment, cooperation and cohesion, while noting mixed findings about synchrony \citep{watsonjones2016}. This perspective directs attention to why people join, persist in and conform to collective activity. Such participation could help sustain the conditions for carrying. Affiliation, agreement on conduct and authorship therefore need separate evidence rather than serving as interchangeable indicators of one process.

Halperin's material-religion account addresses how divine presence becomes tangible through cultivated encounters, using Meyer's concept of sensational forms \citep[pp.~306-307, 313-315]{halperin2026}. It locates recognisable movement within practices through which the assembled rath is encountered as a deity. These accounts can thus contribute differently: attention to correct performance concerns how an act is carried out; group functions concern participation and relationships; cultivated presence concerns the object and significance of that participation. The proposed synthesis is that these relations can organise the physical task and its interpretation without uniquely determining the local force sequence.

Music belongs to this setting as well. Sutherland describes drumming as organising the embodiment, voicing and movement of divine power during consultation \citep[p.~103]{sutherland2006}. That account concerns a ritual relation among participants, not a measured relation between musical beat, gait phase and roll. The numerical inputs below contain no musical recording or entrainment mechanism. Their conditional response shows why a claim about musical effects would need to specify which supports change, in what way, and within what instructed or expected action.

\section{Methods}

The study combines selected published ethnography, experimental research and synthetic forward mechanics. It includes no new interviews, participant experiments or procession measurements. Sources were selected for descriptions of handling, learning, movement, communication and interpretation; this is not a systematic review. The comparison identifies what each passage can establish: a narrated event, an account of practice, a reported sensation or a task-specific experimental result. It then distinguishes the present interpretive proposal from the source author's explanation and from the numerical consequences of assumed inputs. The selected passages do not form an exhaustive or independent sample of ritual episodes. Kullu accounts remain distinct from Rohru, Tukpa and other regional settings. Relevant methods in the author-deposited full texts of the object-manipulation papers were inspected; source versions and locators are recorded in Supplementary Section~S1.

Figure~1 selects the Kullu chair-type rath associated with Ha\d{d}imb\={a} Dev\={i} and Manu \d{R}\d{s}i. Its original geometry follows the structural features documented by Halperin and Luchesi, without tracing their photographs \citep[pp.~14-15]{halperin2016}; \citep[Fig.~3, p.~71]{luchesi2006}. Two simplified people stand at the front and rear between the longitudinal poles. Their number and posture are drawing choices; obscured contacts and load sharing cannot be recovered from these images.

\mainfigure{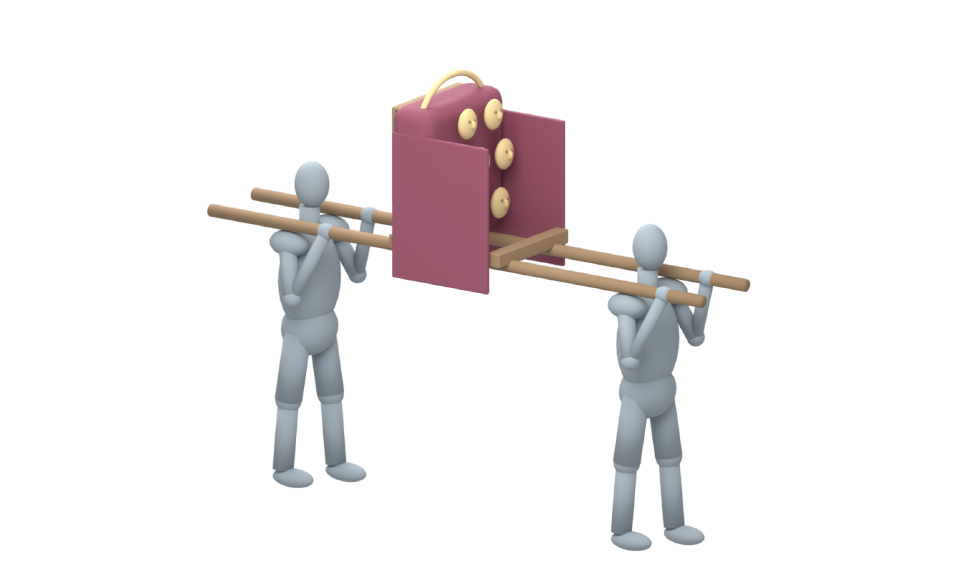}{.82}{\captionone}

Two resultants do not mean two people. Multiple vertical contacts on the same lateral line can be combined while preserving force and roll moment; contacts at differing positions require a force-weighted moment arm. A spring-damper law requires more. If contacts on one side share the object's contact height, remain on their active linear branches and move together, their stiffnesses and damping coefficients add. Proportional damping is not required in this common-motion case. With differing support motions, it is a sufficient condition for one stiffness-weighted effective height to represent both terms. Asynchronous motion therefore does not automatically invalidate the reduction. Contact opening, changing body-contact heights or general damping ratios require additional treatment, as derived in Supplementary Section~S2. These conditions define the reduction used below.

The reference body has two degrees of freedom: heave, or vertical translation, and roll about its fore-aft axis. Each effective support can push upward but cannot pull. Its prescribed height is an input; body pose and forces are solved outputs (Figure~2). The reference mass is 40~kg, roll inertia 8~kg~m$^2$, lateral half-spacing 0.50~m and COM offset 0.15~m above the contact line. Each support has stiffness 8000~N/m and damping 250~N~s/m. All are assumed. The 24.525-mm static compression is an effective coordinate difference, not measured tissue deformation or pad thickness.

This reference isolates how vertical support changes combine into upward motion and sideways tilt. The extension below adds pitch, the forward-backward tilt that front-rear differences can drive. Both models hold horizontal translation and yaw outside the calculation and omit friction and pole bending. Supports remain horizontally registered under the body contacts; the models do not calculate how bearers maintain that arrangement. Horizontal forces and bending could change the calculated response, and their contributions have not been bounded for a real rath. The calculations supply conditional examples rather than an established approximation to a procession.

\mainfigure{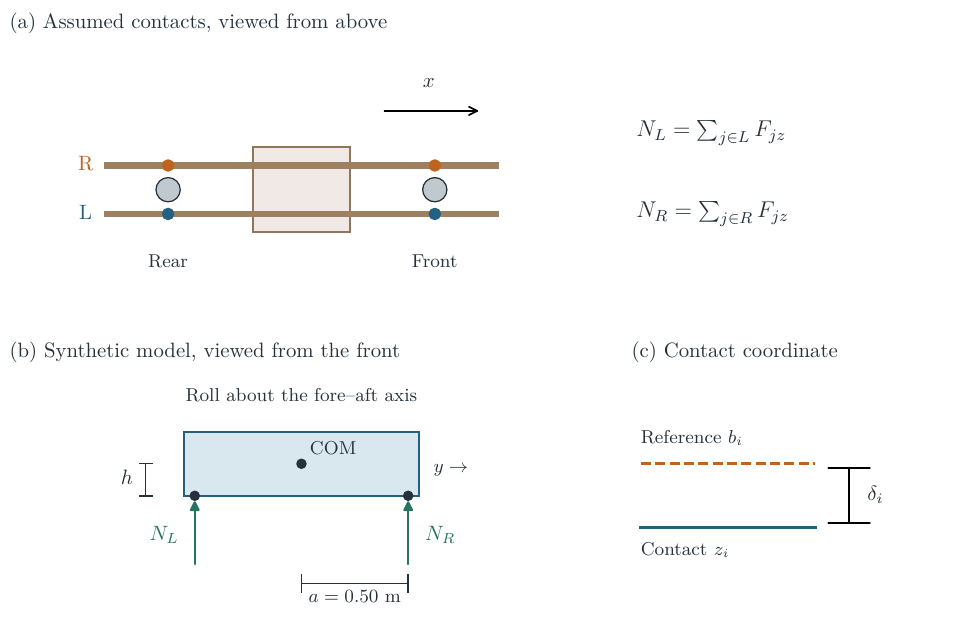}{1.0}{\captiontwo}

Matched cases prescribe equal 4-mm amplitudes at 1.5~Hz and phase differences of zero or 60 degrees, following a two-second ramp. RMS statistics use 4-12~s inclusive; heave is measured about the sample mean and roll about zero. The reference design comprises 94 integrations, including a frequency-phase map, asymmetry, damping and time-step comparisons. A separate structural analysis compares eight parameter sets at both phase differences, giving 16 additional integrations. It changes one assumed quantity at a time, keeps damping fixed and starts each case at its own static equilibrium with zero roll. This design isolates structural effects under a fixed driving input.

\subsection{Representing walking through shoulder motion}

Walking contributes a structured input to a carried object. Trunk-motion research distinguishes two vertical oscillations from one frontal-plane oscillation within a stride \citep{thorstensson1984}. A pole-carrying experiment tests step frequencies including 1.83, 2.00 and 2.33~Hz \citep{castillo2014}. These findings motivate the extension's periodic structure and cadence contrasts; they do not provide measured rath-carrier shoulder trajectories. Paired hand-carrying research also shows changes in gait timing when people share a load \citep{lanini2017}. The present prescribed inputs cannot predict that adaptation.

Two carriers are represented by front and rear pairs of shoulder supports. Each pair has vertical midpoint motion twice per stride and torso tilt once per stride, making one shoulder higher than the other. The body responds through heave, roll and pitch, with four separate compression-only forces. Front-rear spacing is assumed to be 1.50~m; each patch has half the reference side stiffness and damping. Supplementary Section~S14 gives the geometry, inertias, equations and provenance. Step timing enters through these shoulder motions; footsteps and their ground forces are not dynamically solved.

The principal comparison uses 2 steps/s, 20-mm vertical amplitude and 1.5-degree torso-tilt amplitude, with front-rear stride phases of 0, 90 and 180 degrees. Four further cases vary carrier inputs, cadence or input size. Amplitudes and waveform phases are illustrative assumptions. Each case is solved with free pitch, with pitch locked but four contacts retained, and with front-rear inputs averaged before the force law into two side resultants. These comparisons distinguish omitted pitch from contact aggregation. The 21 comparisons last 24~s and use 8-24~s for summaries; 14 further integrations check reference recovery and numerical accuracy, separately from the 94 reference and 16 structural runs.

\section{Findings: what timing and structure determine}

The basic distinction can be stated without equations. If the two supports rise together, their shared upward motion drives the body vertically. If one support rises relative to the other, their unequal loads can create a turning moment. The response also depends on the body's inertia and the support properties, so an input difference does not prescribe a particular angle. In the symmetric small-motion limit, the average input and the difference between inputs separate these two effects.

For left and right deviations $\beta_L=A\sin\omega t$ and $\beta_R=A\sin(\omega t+\Delta)$, the common and differential components are
\begin{align}
\beta_+&=(\beta_L+\beta_R)/2=A\cos(\Delta/2)\sin(\omega t+\Delta/2),\\
\beta_-&=(\beta_R-\beta_L)/2=A\sin(\Delta/2)\cos(\omega t+\Delta/2).
\end{align}
In the symmetric small-angle limit, the common component drives heave and the differential component drives roll. Inertia, stiffness and damping set the response amplitudes and phases. This is an established modal decomposition, used here to make a proposed relation between support timing and object movement explicit.

With zero offset the reference body heaves at 3.543~mm RMS with negligible steady roll. A 60-degree offset gives 3.069~mm heave RMS and 0.1966-degree roll RMS (Table~\ref{tab:matched}; Figure~3). The input offset is 0.111~s at 1.5~Hz: one sixth of the imposed cycle. It is a difference between prescribed support-height signals, not a measured neural delay or a simulated difference between footsteps. RMS describes the typical size of the computed variation over the stated window; it is not the maximum tilt. The heave amplitude ratio predicted by the linear limit is $\cos30^\circ\simeq0.86603$; a harmonic fit to the saved responses gives 0.866025. The result does not require a newly identified nonlinear mechanism.

\begin{table}[!htbp]
\centering\small
\caption{Reference periodic responses over 4-12~s. All cases maintain contact.}
\label{tab:matched}
\begin{tabular}{@{}lrl@{}}
\toprule Input condition & Heave RMS (mm) & Roll RMS (degrees)\\\midrule
0-degree offset & 3.543 & negligible steady roll\\
60-degree offset & 3.069 & 0.1966\\
60-degree offset with load yielding & 3.052 & 0.1957\\
\bottomrule
\end{tabular}
\end{table}

\mainfigure{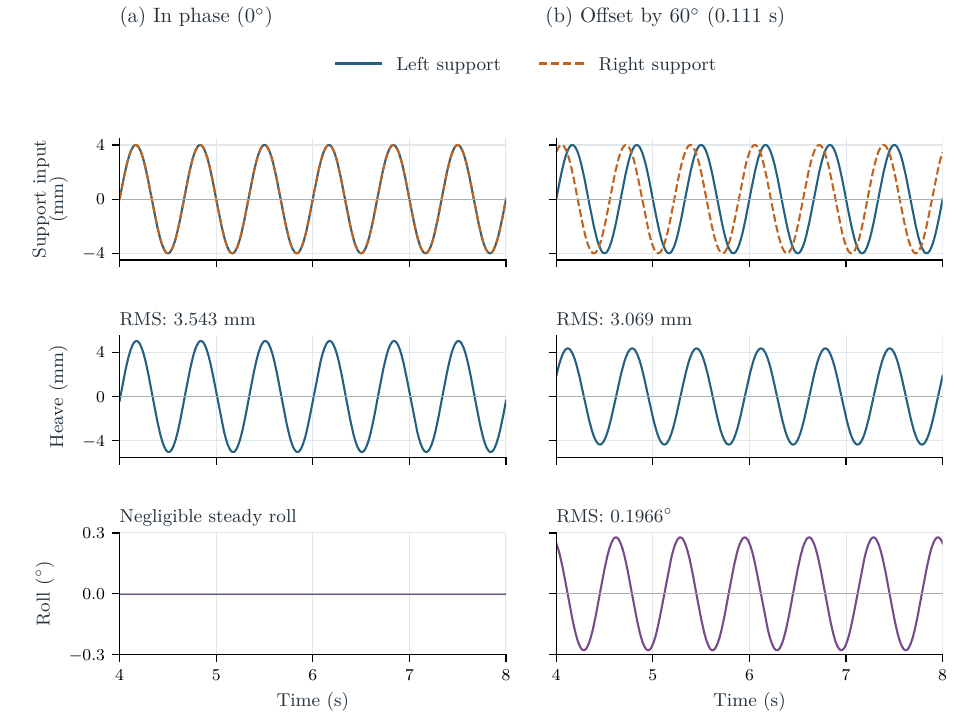}{1.0}{\captionthree}

The 63-cell reference frequency-phase map also separates the largest heave from the largest roll: 5.48~mm RMS at 3~Hz and zero offset, compared with 0.579-degree roll RMS at 3~Hz and 180 degrees. Contacts remain closed.

Case D adds one bounded load-yielding rule: a filtered above-static load commands a downward support adjustment. This is a prescribed feedback algorithm; calling it yielding does not establish mechanical passivity or identify a human reflex. It changes the reported response metrics by less than 1\% and does not account for the main response; roll RMS changes by $-0.446\%$ and RMS force variation by $-0.524\%$. The result limits that rule at the selected settings. It places no general bound on the effects of anticipation, visual guidance, muscle co-contraction or other human feedback mechanisms, which were not represented or compared.

Structural assumptions have much larger effects. The restoring roll stiffness is $K_\theta=2ka^2-Mgh$, so the ratio $\rho=Mgh/(2ka^2)$ measures the gravitational contribution relative to the elastic one; passive stability requires $\rho<1$. Holding other parameters fixed while reducing $a$ from 0.50 to 0.20~m changes $\rho$ from 0.0147 to 0.0920 and lowers the undamped roll frequency from 3.53 to 1.36~Hz. At the same 1.5-Hz, 60-degree drive, roll RMS rises from 0.1966 to 1.183 degrees and peak force from 1.044 to 1.329 times the static force per side. This sixfold roll increase occurs with positive contact throughout. It is a hypothetical spacing change, not a measurement inferred from the drawing.

Across the eight configurations, offset-case roll spans 0.139-1.183 degrees RMS, whereas equal-phase inputs from the symmetric equilibrium produce no roll. The modal distinction survives these particular perturbations; the amplitudes do not remain fixed. Moreover, doubling inertia leaves $\rho$ unchanged while shifting frequency and response. Stability margin alone is therefore not a response predictor: drive frequency, inertia and damping also matter. Supplementary Section~S6 gives the complete compact comparison. Numerical convergence estimates accuracy for a chosen configuration; it cannot remove uncertainty about which configuration represents a real apparatus.

The appropriate synchrony claim is consequently narrow: without specifying support geometry and the motion component being measured, greater timing agreement need not imply greater rocking. Within this equal-amplitude model, phase difference is informative when the other parameters are fixed. The cited ritual studies do not assert that synchrony necessarily increases roll or divine attribution; this calculation specifies a physical requirement for a proposed explanation rather than refuting such a literature-wide position.

\subsection{Walking, pitch and loads that rocking can hide}

Walking-related timing changes which parts of the body move. Matched front and rear strides produce heave and roll. At a 90-degree stride offset, the vertical midpoint motions oppose: heave nearly cancels, while pitch reaches 1.43 degrees RMS alongside 0.825-degree roll (Table~\ref{tab:walking}; Figure~4a,b). This offset is a quarter-stride, or 0.25~s at this cadence; it differs from the reference offset between pole-side inputs.

For the same G90 inputs, the largest patch force is 160.3~N with free pitch and 213.3~N with pitch locked. Free pitch maintains all contacts; the locked case transmits zero force in 11.7\% of contact-time samples. The two-resultant reduction reports closed contact but cannot represent pitch or individual loads. Constraining pitch can therefore misrepresent how the body accommodates a front-rear mismatch.

\begin{table}[!htbp]
\centering\small
\caption{Four-contact, free-pitch responses to walking-informed inputs over 8-24~s. $H$, $R$ and $P$ denote heave, roll and pitch RMS; angular zeros denote values below $10^{-10}$ degrees. Peak patch load is divided by the 98.1-N static patch load; zero force is the fraction of all contact-time samples. Values are synthetic, not typical carrier measurements.}
\label{tab:walking}
\begin{tabular}{@{}lrrrrr@{}}
\toprule Case & $H$ (mm) & $R$ (deg) & $P$ (deg) & Peak / static & Zero force (\%)\\\midrule
Matched strides (G0) & 21.10 & 1.169 & 0 & 1.54 & 0\\
90-degree offset (G90) & 0.046 & 0.825 & 1.427 & 1.63 & 0\\
180-degree offset (G180) & 21.07 & 0 & 0 & 2.00 & 0\\
Varying carriers (Gvar) & 15.37 & 0.386 & 0.778 & 1.96 & 0\\
Larger-input stress case & 48.7 & 0 & 0 & 3.97 & 50.4\\
\bottomrule
\end{tabular}
\end{table}

At 180-degree stride separation, vertical motion repeats twice per stride: the carriers bounce together while their torso tilts oppose. Diagonal load changes cancel the turning moments. Roll and pitch remain negligible, and the reduction recovers heave and side-force sums while hiding individual loads (Figure~4c). The largest patch force is 196.7~N; dividing the reduced side force equally gives a maximum of only 145.1~N. That allocation does not predict each shoulder's force. Little visible rocking can coexist with substantial redistribution of support.

\mainfigure{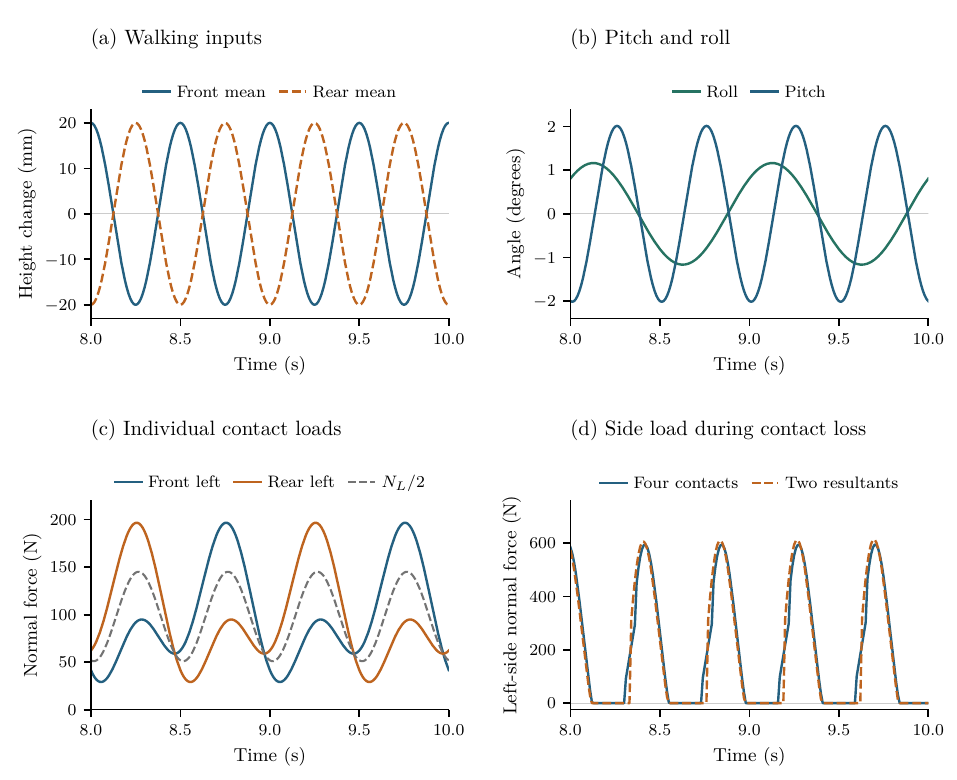}{1.0}{\captionfour}

In the larger-input case, 35-mm vertical amplitude and 3-degree torso tilt at 2.33 steps/s produce repeated contact opening. Averaging inputs before the force law changes the aggregate response: heave is 48.7 rather than 51.5~mm RMS, and peak side force is about 3.04 rather than 3.26 times static side load (Figure~4d). Refined and independent integrations support these rounded summaries, without establishing pointwise force convergence at every recontact. Across the G90 cadence contrasts, pitch ranges from 1.36 to 1.55 degrees RMS. The cadence comparison quantifies how input rate changes the response within the same mechanical system.

\section{Discussion: contributing while following}

\subsection{Contribution without sole control}

An adjustment made in response to a shared load also changes the support it receives. The model makes the second part of this relation explicit: changing one input can affect a motion jointly determined by both sides. The prescribed height signals can represent inputs supplied under instruction, learned preparation or reactive correction. Thus a forward mechanical calculation constrains the relation between inputs and motion without selecting their psychological origin.

The walking extension sharpens this distinction. A nearly upright object need not mean that each carrier experiences an unchanging load. In the diagonal-load example, contact forces vary while their turning effects cancel. Conversely, permitting the body to pitch can accommodate an input difference that causes unloading when pitch is prohibited. These are reasons to measure individual support as well as visible movement when investigating local contribution. The comparison identifies individual load sharing as an essential link between the visible action and the work of its bearers.

This distinction gives the ethnographic comparison a positive explanatory use. The shoulder account makes responsive contribution intelligible; the upright case changes the controlled quantity from rocking to stability; rehearsal concerns a recognisable repertoire; and the return narrative assigns an episode a purpose. A person can intentionally maintain footing and still describe the deity as leading. Such an account need not deny every local adjustment, and acknowledging adjustment does not establish conscious control of the whole episode.

Experience of control requires its own evidence. In a shared-object task, Dewey and colleagues examined judgments of individual and joint influence; related work distinguishes implicit timing measures from explicit agency judgments \citep{deweypacherie2014,deweyknoblich2014}. Reviews likewise distinguish self, partner and collective agency, and the cues used in agency judgments \citep{loehr2022,synofzik2008}. These distinctions guide questions about carriers' reports but do not translate religious authorship into a laboratory score. An account of divine initiative may express a compelling experience, a recognised obligation, or both; neither a load trace nor a measure of group affiliation identifies which is at issue.

Comparison across practices clarifies what counts as relevant evidence. Dubisch's icon bearers at Tinos slow to accommodate kneeling pilgrims: a specific adjustment serves participation in worship \citep[pp.~26-27]{dubisch1995}. The example connects practical motion to an established ritual action, but does not provide a matched contrast in agency attribution with a Himalayan rath. Konvalinka and colleagues find arousal patterned by relationships between fire-walkers and spectators \citep{konvalinka2011}. Their result illustrates why participation cannot be treated as exposure to the same stimulus alone. It does not measure support timing or religious authorship. These comparisons motivate separate observations of movement, relationships and attribution; they do not establish a common causal mechanism across the traditions.

\subsection{Reading the upright-rath account with the load comparison}

Halperin records Rajender of Dhungri explaining that some large raths, including Bijl\={i} Mah\={a}dev's, remain upright because swinging would endanger their carriers, while retaining the deity's power and distinctive style \citep[pp.~321-322]{halperin2026}. The account directs attention to a different achievement from a conspicuous bow: maintaining a posture can itself be part of how a deity is recognised and carried.

The synthetic G180 case addresses an inference that this account brings into focus. Opposing diagonal load changes cancel their turning effects, leaving negligible roll and pitch despite substantial changes in individual support. The example demonstrates that an upright orientation does not require unchanging individual loads. It does not estimate the forces beneath Bijl\={i} Mah\={a}dev's rath or show that its carriers use this particular phase relation. Equally, the example does not make variable loads necessary: other inputs or active corrections could maintain the same posture.

The connection therefore has three parts. The source identifies upright carrying as meaningful and practically consequential. The calculation provides a counterexample to inferring constant individual support from little angular movement. The proposed empirical consequence is to record individual loads and adjustments alongside orientation when examining that practice. If those loads were stable in an observed episode, this particular redistribution example would not explain its support; if they varied, their timing and the available cues would still be needed to distinguish guidance, anticipation and correction. This comparison yields a practical measurement strategy that connects the meaningful posture to the distribution of support.

\subsection{What would make the explanation more specific}

The theoretical comparison offers a way to connect evidence to a claim. If the claim concerns local mechanical influence, it requires records of support motion or load together with object motion. An apparatus or replica assessment would first need mass, contact positions and a defensible estimate of inertia and compliance. Simultaneous records of foot timing, shoulder motion, front-rear tilt and individual loads would test the extension's prescribed coupling; horizontal movement and pole deflection would test remaining omissions. A match of one roll trace would be insufficient if different inputs or parameter choices could produce it.

If the claim concerns learned anticipation, the relevant evidence changes. Preparation must be related to handling history and the cues available before the load change, with current instructions and partner behaviour considered. If it concerns felt control or the purpose of an episode, participants' accounts are indispensable. Expectations before an event, descriptions of immediate sensation and later narratives answer related but distinct questions. These comparisons need not all be collected in one study. Each should connect the evidence to a particular proposed cause. A temporal lead by a speaker or an experienced bearer could reflect shared anticipation of a known event, so order alone would not establish causal influence. Disagreement, unanticipated load changes and differences in who could perceive a cue could help distinguish the explanations when documented in context.

For a route refusal or a return, additional evidence would have to concern actual progress, stopping and turning, together with intended destinations and situated accounts of the decision. Pitch is a change of orientation, not progress along a route. Rishi's narrative establishes an attributed purpose for a return. Connecting that purpose to the initiating actions requires a record of the route, communication and support through the turn.

\subsection{Scope and implications}

The study combines a purposive source comparison with synthetic mechanics. It reconstructs no particular procession or route decision; carrier intentions, handling experience and effects of familiarity or ritual speech on control and divine attribution remain unmeasured. Anticipation, leadership and attention are therefore explanatory hypotheses for carrying. Narratives, observations and experimental tasks retain their distinct evidential roles; historical and regional descriptions are not pooled as one event record. The source authors' explanations of presence and sovereignty remain essential to interpreting their material.

The contact derivation supplies sufficient conditions for a reduction, not evidence that the contacts in a procession meet them. The numerical supports are prescribed and do not learn or adapt their gait to the load. The walking extension adds pitch and separate contacts, while retaining rigid geometry, vertical forces and constrained horizontal motion and yaw. Its within-stride waveform shapes and relative timing are assumptions, not fitted shoulder kinematics. The parameter changes are independent mathematical perturbations, although real mass, inertia, dimensions and contact properties could covary. The analysis neither spans a measured population nor tests all stable parameter combinations. Its computed millimetres and degrees belong to the declared synthetic system. Faithful rendering does not calibrate the illustration or validate a social explanation.

Regional scope also concerns different kinds of claim. The illustrated Kullu chair-type form is not a template for Rohru or Mah\={a}s\={u} apparatuses, and none of the numerical values is an estimate for those traditions. The questions about preparation, influence and interpretation can nevertheless guide comparison elsewhere. Their usefulness would have to be assessed against each practice's own material arrangements and accounts of action. Analytical portability is a proposal for comparison, not empirical generalisation across Himalayan traditions.

The specific advance is to locate an apparent conflict between carrying and following at the appropriate scale. Knowing where an episode is going, preparing for local load, influencing its motion and attributing its purpose are separable achievements. Evidence for one does not settle the others. The mechanical examples show how prescribed local inputs contribute to shared motion; the source contrasts distinguish the roles of contribution and following in rehearsal, posture, consultation and narration. This makes instruction, anticipation, correction and interpretation distinguishable without requiring that only one operate in a ritual event.

Returning to Rishi's story, explaining how the rath was supported would leave open why the return to the seller was understood as a lesson. Explaining that interpretation would leave open the sequence of local adjustments. The paper connects those questions through the organisation of a task, the distribution of practical influence and the interpretation of its outcome. A person can contribute to sustaining a movement, respond to demands they did not individually choose, and understand the event through a deity's initiative. The resulting account brings technical competence, authorised initiative and experienced following into a shared explanation of participation. It identifies the task, the local contributions that sustain it and the attribution through which the episode acquires purpose.

\subsection*{Code  and data availability}
All research code, simulation data, numerical outputs and reproduction materials are available at \url{https://github.com/nalin-dhiman/palanquin-coupled-oscillator-dynamics}.
\subsection*{Ethics and informed consent}
This study uses published scholarship and synthetic numerical data. It involved no participant recruitment, interviews, interventions or collection of identifiable personal data. No participant consent was sought or required for these activities. No institutional ethics approval is claimed.

\section*{Acknowledgements}
The author acknowledges the published fieldwork and scholarship of E. Halperin, P. Sutherland, D. Berti, B. Luchesi, J. Dubisch and the other researchers cited in this article, whose documented accounts make the present synthesis possible. The author also acknowledges the communities and participants represented in those published studies, and the developers and maintainers of Python, NumPy, SciPy, Matplotlib, Blender ecosystem used for computation and visualisation.

\section*{Author contributions}
N.D. conceived the study, developed the methodology and software, conducted the source analysis and numerical investigations, validated the results, prepared the visualisations, and wrote and edited the manuscript. N.D. is the sole author and is responsible for all aspects of the work.

\section*{Funding}
This research received no funding.

\section*{Additional information}
\noindent\textbf{Competing interests.} The author declares no competing interests.

\begingroup\small
\setlength{\bibsep}{0pt}
\setlength{\parskip}{0pt}

\endgroup

\clearpage
\includepdf[
    pages=-,
    pagecommand={},
    fitpaper=true
]{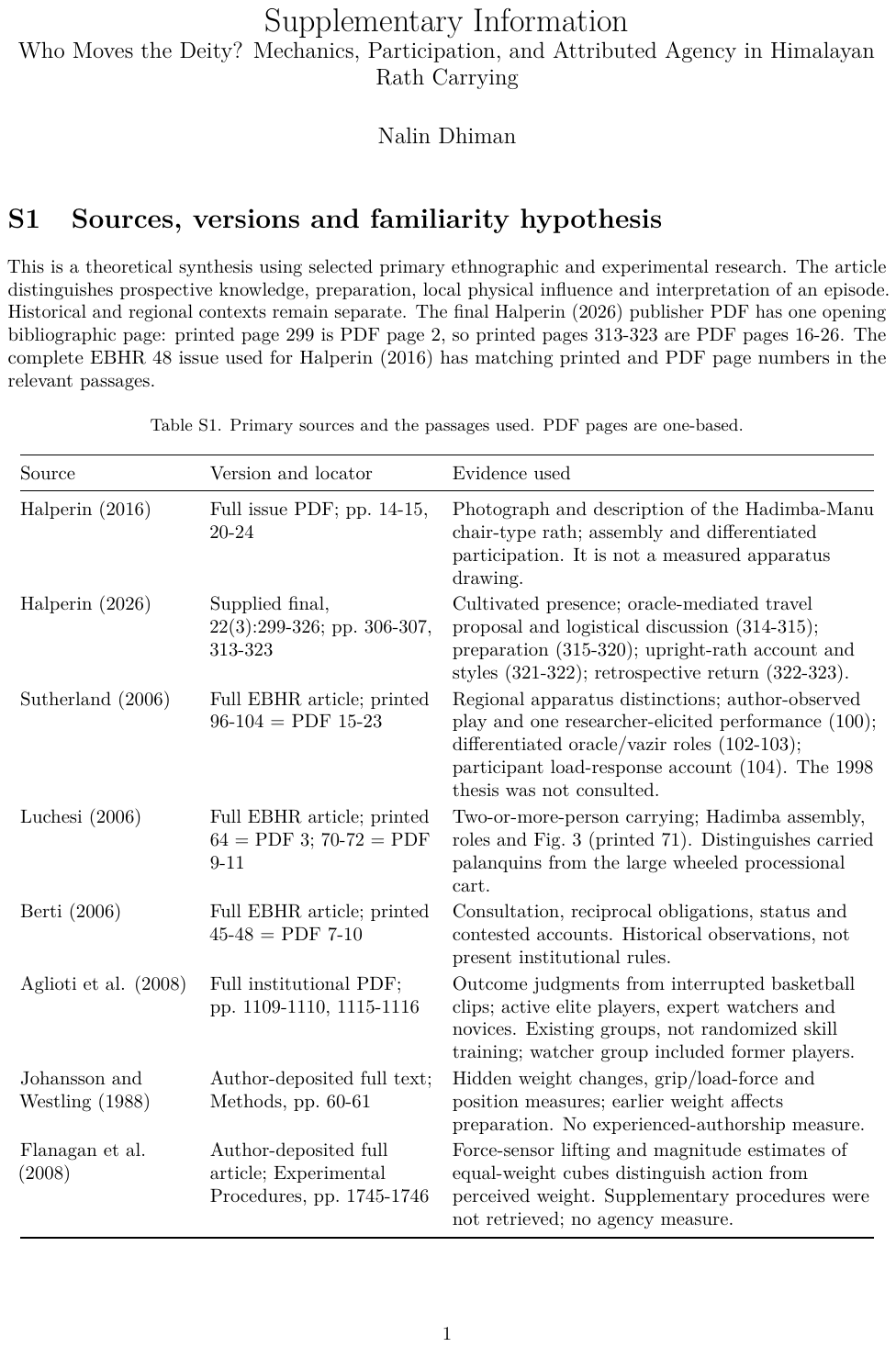}


\begin{thebibliography}{99}
\bibitem{halperin2026}
Halperin, E. Gods on the move: Ritual experience and the formation of dynamic divine presence in the Western Himalaya. \emph{Mater. Relig.} \textbf{22}, 299-326 (2026). \url{https://doi.org/10.1080/17432200.2026.2688669}

\bibitem{halperin2016}
Halperin, E. A vehicle for agency: Rath rituals and the construction of Himalayan devtas as complex agents. \emph{European Bulletin of Himalayan Research} \textbf{48}, 5-42 (2016).

\bibitem{luchesi2006}
Luchesi, B. Fighting enemies and protecting territory: Deities as local rulers in Kullu, Himachal Pradesh. \emph{European Bulletin of Himalayan Research} \textbf{29-30}, 62-81 (2006).

\bibitem{meyer2006}
Meyer, B. \emph{Religious Sensations: Why Media, Aesthetics and Power Matter in the Study of Contemporary Religion} (Vrije Universiteit, 2006).

\bibitem{sutherland2006}
Sutherland, P. T(r)opologies of rule (R\={a}j): Ritual sovereignty and theistic subjection. \emph{European Bulletin of Himalayan Research} \textbf{29-30}, 82-119 (2006).

\bibitem{berti2006}
Berti, D. Ritual kingship, divine bureaucracy and electoral politics in Kullu. \emph{European Bulletin of Himalayan Research} \textbf{29-30}, 39-61 (2006).

\bibitem{loehr2022}
Loehr, J. D. The sense of agency in joint action: An integrative review. \emph{Psychon. Bull. Rev.} \textbf{29}, 1089-1117 (2022). \url{https://doi.org/10.3758/s13423-021-02051-3}

\bibitem{johansson1988}
Johansson, R. S. \& Westling, G. Coordinated isometric muscle commands adequately and erroneously programmed for the weight during lifting task with precision grip. \emph{Exp. Brain Res.} \textbf{71}, 59-71 (1988). \url{https://doi.org/10.1007/BF00247522}

\bibitem{flanagan2008}
Flanagan, J. R., Bittner, J. P. \& Johansson, R. S. Experience can change distinct size-weight priors engaged in lifting objects and judging their weights. \emph{Curr. Biol.} \textbf{18}, 1742-1747 (2008). \url{https://doi.org/10.1016/j.cub.2008.09.042}

\bibitem{aglioti2008}
Aglioti, S. M., Cesari, P., Romani, M. \& Urgesi, C. Action anticipation and motor resonance in elite basketball players. \emph{Nat. Neurosci.} \textbf{11}, 1109-1116 (2008). \url{https://doi.org/10.1038/nn.2182}

\bibitem{pruszynski2011}
Pruszynski, J. A., Kurtzer, I. \& Scott, S. H. The long-latency reflex is composed of at least two functionally independent processes. \emph{J. Neurophysiol.} \textbf{106}, 449-459 (2011). \url{https://doi.org/10.1152/jn.01052.2010}

\bibitem{yang2011}
Yang, L., Michaels, J. A., Pruszynski, J. A. \& Scott, S. H. Rapid motor responses quickly integrate visuospatial task constraints. \emph{Exp. Brain Res.} \textbf{211}, 231-242 (2011). \url{https://doi.org/10.1007/s00221-011-2674-3}

\bibitem{boyer2006}
Boyer, P. \& Li\'enard, P. Why ritualized behavior? Precaution systems and action parsing in developmental, pathological and cultural rituals. \emph{Behav. Brain Sci.} \textbf{29}, 595-613 (2006). \url{https://doi.org/10.1017/S0140525X06009332}

\bibitem{watsonjones2016}
Watson-Jones, R. E. \& Legare, C. H. The social functions of group rituals. \emph{Curr. Dir. Psychol. Sci.} \textbf{25}, 42-46 (2016). \url{https://doi.org/10.1177/0963721415618486}

\bibitem{thorstensson1984}
Thorstensson, A., Nilsson, J., Carlson, H. \& Zomlefer, M. R. Trunk movements in human locomotion. \emph{Acta Physiol. Scand.} \textbf{121}, 9-22 (1984). \url{https://doi.org/10.1111/j.1748-1716.1984.tb10452.x}

\bibitem{castillo2014}
Castillo, E. R., Lieberman, G. M., McCarty, L. S. \& Lieberman, D. E. Effects of pole compliance and step frequency on the biomechanics and economy of pole carrying during human walking. \emph{J. Appl. Physiol.} \textbf{117}, 507-517 (2014). \url{https://doi.org/10.1152/japplphysiol.00119.2014}

\bibitem{lanini2017}
Lanini, J., Duburcq, A., Razavi, H., Le Goff, C. G. \& Ijspeert, A. J. Interactive locomotion: Investigation and modeling of physically-paired humans while walking. \emph{PLOS ONE} \textbf{12}, e0179989 (2017). \url{https://doi.org/10.1371/journal.pone.0179989}

\bibitem{deweypacherie2014}
Dewey, J. A., Pacherie, E. \& Knoblich, G. The phenomenology of controlling a moving object with another person. \emph{Cognition} \textbf{132}, 383-397 (2014). \url{https://doi.org/10.1016/j.cognition.2014.05.002}

\bibitem{deweyknoblich2014}
Dewey, J. A. \& Knoblich, G. Do implicit and explicit measures of the sense of agency measure the same thing? \emph{PLOS ONE} \textbf{9}, e110118 (2014). \url{https://doi.org/10.1371/journal.pone.0110118}

\bibitem{synofzik2008}
Synofzik, M., Vosgerau, G. \& Newen, A. Beyond the comparator model: A multifactorial two-step account of agency. \emph{Conscious. Cogn.} \textbf{17}, 219-239 (2008). \url{https://doi.org/10.1016/j.concog.2007.03.010}

\bibitem{dubisch1995}
Dubisch, J. \emph{In a Different Place: Pilgrimage, Gender, and Politics at a Greek Island Shrine} (Princeton University Press, 1995).

\bibitem{konvalinka2011}
Konvalinka, I. et~al. Synchronized arousal between performers and related spectators in a fire-walking ritual. \emph{Proc. Natl Acad. Sci. USA} \textbf{108}, 8514-8519 (2011). \url{https://doi.org/10.1073/pnas.1016955108}
\end{thebibliography}
\end{document}


\maketitle

\section{Sources, versions and familiarity hypothesis}

This is a theoretical synthesis using selected primary ethnographic and experimental research.  The article distinguishes prospective knowledge, preparation, local physical influence and interpretation of an episode. Historical and regional contexts remain separate. The final Halperin (2026) publisher PDF has one opening bibliographic page: printed page 299 is PDF page 2, so printed pages 313-323 are PDF pages 16-26. The complete EBHR 48 issue used for Halperin (2016) has matching printed and PDF page numbers in the relevant passages.

\begin{longtable}{@{}>{\raggedright\arraybackslash}p{.19\linewidth}>{\raggedright\arraybackslash}p{.25\linewidth}>{\raggedright\arraybackslash}p{.48\linewidth}@{}}
\caption{Primary sources and the passages used. PDF pages are one-based.}\\
\toprule Source & Version and locator & Evidence used\\\midrule\endfirsthead
\toprule Source & Version and locator & Evidence used\\\midrule\endhead
Halperin (2016) & Full issue PDF; pp.~14-15, 20-24 & Photograph and description of the Hadimba-Manu chair-type rath; assembly and differentiated participation. It is not a measured apparatus drawing.\\
Halperin (2026) & Supplied final, 22(3):299-326; pp.~306-307, 313-323 & Cultivated presence; oracle-mediated travel proposal and logistical discussion (314-315); preparation (315-320); upright-rath account and styles (321-322); retrospective return (322-323).\\
Sutherland (2006) & Full EBHR article; printed 96-104 = PDF 15-23 & Regional apparatus distinctions; author-observed play and one researcher-elicited performance (100); differentiated oracle/vazir roles (102-103); participant load-response account (104). The 1998 thesis was not consulted.\\
Luchesi (2006) & Full EBHR article; printed 64 = PDF 3; 70-72 = PDF 9-11 & Two-or-more-person carrying; Hadimba assembly, roles and Fig.~3 (printed 71). Distinguishes carried palanquins from the large wheeled processional cart.\\
Berti (2006) & Full EBHR article; printed 45-48 = PDF 7-10 & Consultation, reciprocal obligations, status and contested accounts. Historical observations, not present institutional rules.\\
Aglioti et al. (2008) & Full institutional PDF; pp.~1109-1110, 1115-1116 & Outcome judgments from interrupted basketball clips; active elite players, expert watchers and novices. Existing groups, not randomized skill training; watcher group included former players.\\
Johansson and Westling (1988) & Author-deposited full text; Methods, pp.~60-61 & Hidden weight changes, grip/load-force and position measures; earlier weight affects preparation. No experienced-authorship measure.\\
Flanagan et al. (2008) & Author-deposited full article; Experimental Procedures, pp.~1745-1746 & Force-sensor lifting and magnitude estimates of equal-weight cubes distinguish action from perceived weight. Supplementary procedures were not retrieved; no agency measure.\\
\bottomrule
\end{longtable}

The familiarity hypothesis concerns preparation beyond that supplied by current cues or instructions. The comparisons in main Table~1 locate the required evidence; none is implemented by the prescribed supports. Carrier histories, experience of control and attribution were not measured. Full main-article methods of Johansson-Westling and Flanagan et al. were inspected in the publicly accessible texts deposited by Roland S.~Johansson on ResearchGate. The former uses a small precision-grip manipulandum, not a shared shoulder load. The latter measures force development and perceived weight, not authorship. Their application to carrying remains a theoretical extension.

Boyer-Li\'enard's target article develops a proposed relation between prescribed performance, attention to action details and concerns about correct performance (pp.~610-611, especially Fig.~4). It distinguishes ritualization from routinization; the main text applies that distinction as a hypothesis about attention, not as evidence of carriers' psychological states. The supplied Dubisch book supports bodily accommodation at Tinos (pp.~26-27); the supplied Konvalinka paper concerns relationship-dependent arousal, not motor synchrony. Watson-Jones and Legare's five-page article was inspected in an author-linked publisher PDF. Its account of membership, commitment, cooperation and cohesion (pp.~42-44) concerns proposed group functions; its discussion of synchrony notes mixed findings. The main text distinguishes these outcomes from authorship rather than treating affiliation as an agency measure. Motor-correction and joint-agency sources in the main bibliography support the task-specific distinctions stated there. Meyer's concept is retained with Halperin's explicit use of it; the present article does not attribute the proposed load-learning psychology to either scholar.

\subsection{Connecting source passages and mechanical claims}

The communication example uses Halperin's recorded conversation with Mahesh (printed pp.~314-315, PDF pp.~17-18). The deity's wish to travel is reported as conveyed through an oracle; the priest describes what the k\={a}rd\={a}r would need to do to seek space. This passage does not establish that the manager approached the administration, that space was secured or that any particular carrier heard the oracle. The analysis concerns the relation between an attributed initiative and an assigned practical responsibility. Sutherland's account of questioning, oracular speech and execution (pp.~102-103) is a separate Rohru/Tukpa comparison, not evidence for the same offices or sequence in Kullu.

The upright-carrying comparison uses Rajender's account in Halperin (printed pp.~321-322, PDF pp.~24-25). It is a statement about particular raths' conduct, danger and divine character, not a measured force record. G180 supplies a synthetic counterexample to constant load inferred from negligible angular motion. No numerical parameter, stride relation or force magnitude is assigned to Bijl\={i} Mah\={a}dev's apparatus. Variable loads, stable loads and active corrective strategies remain empirical alternatives for that practice.

The theoretical synthesis links three evidential operations: identifying what action matters in a source, establishing a conditional material consequence, and specifying observations that could distinguish explanations in an episode. 

\section{Contextual geometry and the resultant reduction}

The original static schematic selects the Kullu chair-type apparatus documented for Hadimba Devi and Manu Rishi. Halperin (2016), printed p.~14, and Luchesi (2006), Fig.~3 on printed p.~71, were inspected as contextual photographs. Halperin's description on p.~15 and practical account in 2026 (pp.~318-320) inform the shoulder-carried arrangement. A wooden chair/frame carries cloth and metal faces; two longitudinal poles project past the load and the shown carriers. The front carrier is visibly between the poles in the primary reference. Rear contacts and other assistance are partly obscured. Two front/rear silhouettes are a declared simplification, not a claim to count all bearers or helpers in a photograph. Hands are contextual and supply no simulated force. The mesh uses original, simplified cloth, face and frame forms rather than a close redraw of protected decoration. It contains no temple idol or borrowed Rohru longhair form.

The contextual mesh uses arbitrary proportional units: pole length 3.20, lateral separation 0.40, front/rear carrier coordinates $x=\pm1.10$ and pole-centre height 1.47. These are drawing choices, not measured metric parameters. No mass, inertia, pole stiffness or COM is inferred from the pictured silhouette. The separately drawn synthetic model retains $2a=1.00$~m and $h=0.15$~m. The multi-angle illustration is static. A separate illustrative animation uses this same mesh as a rigid graphical attachment to the synthetic model pose. Its appearance does not supply the missing mass, inertia, COM or contact calibration; it is not a physical reconstruction of the illustrated apparatus.

Let $F_{jz}$ be the vertical force at actual contact $j$, with world-space lateral coordinate $y_j$ relative to the COM. Grouping candidate contacts by pole side gives
\begin{equation}
N_L=\sum_{j\in L}F_{jz},\qquad N_R=\sum_{j\in R}F_{jz},\qquad
\tau_x=\sum_j y_jF_{jz}.
\end{equation}
Preserving total vertical load is insufficient unless the moment is also preserved. Where all contacts on side $i$ have the same $y_i$, the moment reduces to $y_LN_L+y_RN_R$. Otherwise the effective position must satisfy
\begin{equation}
y_i^{\mathrm{eff}}=\frac{\sum_{j\in i}y_jF_{jz}}{N_i}\quad(N_i\ne0).
\end{equation}
It may move with load redistribution. For rigidly attached points sharing body-frame $(s_i,-h)$ in the lateral plane, $y_i=s_i\cos\theta+h\sin\theta$ even when their longitudinal coordinates differ. Those longitudinal coordinates instead enter the pitch moment $\tau_y=-\sum_j x_jF_{jz}$ for vertical forces. Pitch must be balanced or constrained separately; it is not represented by the roll equation. Horizontal forces, couples and flexible pole deformation would add terms or degrees of freedom.

\subsection{A sufficient constitutive reduction}

Force aggregation alone does not establish a spring-damper law. Consider patches $j$ on one side $i$. Assume that their body-contact heights and velocities are the same, $z_i$ and $\dot z_i$, their coefficients $k_{ij}>0$ and $c_{ij}\ge0$ are constant, and every patch remains compressed with positive raw force. On this active linear branch,
\begin{align}
N_i&=\sum_j\{k_{ij}(b_{ij}-z_i)+c_{ij}(\dot b_{ij}-\dot z_i)\},\\
K_i&=\sum_j k_{ij},\qquad C_i=\sum_j c_{ij}.
\end{align}
If all $b_{ij}=b_i$, it follows directly that
\begin{equation}
N_i=K_i(b_i-z_i)+C_i(\dot b_i-\dot z_i).
\end{equation}
No proportionality between the patch coefficients is needed in this common-motion case.

For differing support motions, define the stiffness-weighted height
\begin{equation}
b_i^{K}=\frac{\sum_j k_{ij}b_{ij}}{K_i}.
\end{equation}
Constant coefficients imply $\dot b_i^K=\sum_jk_{ij}\dot b_{ij}/K_i$. If $c_{ij}=\lambda_i k_{ij}$ for every patch on the side, then $C_i=\lambda_i K_i$ and
\begin{equation}
N_i=K_i(b_i^K-z_i)+C_i(\dot b_i^K-\dot z_i).
\end{equation}
Thus one effective height represents both elastic and viscous terms under this additional sufficient condition. With general damping ratios the exact expression instead uses separately weighted inputs,
\begin{equation}
N_i=K_i(b_i^K-z_i)+C_i(\dot b_i^C-\dot z_i),\qquad
b_i^C=\frac{\sum_jc_{ij}b_{ij}}{C_i}\quad(C_i>0).
\end{equation}
These need not be the same height. A dynamically defined effective input might still be constructed, but it would not be the fixed stiffness-weighted input assumed above.

The active-branch restriction matters. Summing separately gated or clipped patch forces generally differs from gating or clipping their aggregate. For example, two undamped patches with $k=1000$~N/m and compressions 0.02 and $-0.01$~m give a total force of 20~N; using their mean compression with stiffness 2000~N/m gives 10~N. Moving contact arms also require the moment calculation above. The two-resultant model is therefore justified for shared body kinematics, appropriate support-motion weighting and active contacts, with pitch balance treated separately. These conditions have not been measured for the illustrated rath. 

\section{Scope and coordinate convention}

Sections S3-S11 derive and audit the reference model in one lateral-vertical plane. Sections S14-S16 add the walking-informed, four-contact extension with pitch. World $x$ points forward, $y$ is lateral and $z$ points upward. In the reference notation, roll $\theta$ is a right-handed rotation about $+x$; positive $\theta$ raises the positive-$y$ end. The rigid core has center of mass position $(0,0,z)$ and roll inertia $I$. Its two effective body-frame contact points are
\begin{equation}
\mathbf r_i^B=(0,s_i,-h),\qquad s_L=-a,\quad s_R=+a.
\end{equation}
They are specified synthetic effective points. The conditional aggregation above explains how vertical resultants may represent multiple contacts; it does not identify this fixed topology with the illustrated people.

The reference model omits fore-aft translation, yaw and pitch, horizontal contact forces, friction, distributed contact moments, pole flexure, decoration, walking bodies and route choice. The ideal vertical supports are assumed to remain horizontally registered under the effective points. This assumption would require a separate frictional or geometric mechanism in a full spatial model; no hidden horizontal constraint moment is included in the present roll balance.

These omissions can affect the calculated coordinates as well as add new ones. Pitch or pole bending can make front and rear body-contact heights differ on the same lateral side, violating S2.1's shared-kinematics condition. A horizontal lateral force $F_{yj}$ at height $z_j-z_{\rm COM}$ adds $-(z_j-z_{\rm COM})F_{yj}$ to the roll moment, beyond the retained vertical-force term $y_jF_{zj}$. No bound on these missing contributions has been measured. The model is a conditional planar example.  Testing that approximation would require contact geometry, compliance and omitted-motion measurements at the intended time and motion scales.

\section{Finite-angle kinematics and equations of motion}

Rotation about $x$ maps a body point's lateral and vertical coordinates by
\begin{equation}
\begin{bmatrix}y_i\\z_i-z\end{bmatrix}=
\begin{bmatrix}\cos\theta&-\sin\theta\\
\sin\theta&\cos\theta\end{bmatrix}
\begin{bmatrix}s_i\\-h\end{bmatrix}.
\end{equation}
Hence
\begin{align}
y_i&=s_i\cos\theta+h\sin\theta,\\
z_i&=z+s_i\sin\theta-h\cos\theta,\\
J_i\equiv\frac{\partial z_i}{\partial\theta}
&=s_i\cos\theta+h\sin\theta=y_i.
\end{align}
An upward force $(0,0,N_i)$ at lateral distance $y_i$ has $x$-moment $y_iN_i$, which verifies the sign of the generalized force. Lagrange's equations, equivalently vertical and roll Newton-Euler balance, give
\begin{align}
M\ddot z&=\sum_iN_i-Mg,\\
I\ddot\theta&=\sum_iJ_iN_i. \label{eq:finite}
\end{align}
For $\theta>0$, the right point rises. Holding support heights fixed then reduces right compression and increases left compression; the resulting moment is negative when the passive stiffness condition below holds. The baseline tests independently check the positive roll-moment sign from unequal support forcing.

\section{Unilateral compliant contact and static preload}

Let $b_i(t)$ be effective support height, including the prescribed drive and any feedback correction. Define compression and compression rate
\begin{align}
\delta_i&=b_i-z_i,\\
\dot\delta_i&=\dot b_i-\dot z-J_i\dot\theta.
\end{align}
The force law is
\begin{equation}
N_i=\begin{cases}
0,&\delta_i\leq0,\\
\max(0,k_i\delta_i+c_i\dot\delta_i),&\delta_i>0.
\end{cases} \label{eq:contact}
\end{equation}
It is unilateral in both gap and force. During rapid decompression, the dashpot cannot pull the body toward a still-compressed support: if the raw Kelvin-Voigt force becomes negative, it is clipped to zero. This is why dissipation cannot everywhere be written as $c\dot\delta^2$ without checking the active branch.

For equal contacts and $\theta=0$, static balance gives
\begin{equation}
N_L=N_R=N_0=\frac{Mg}{2},\qquad
\delta_0=\frac{Mg}{2k}.
\end{equation}
With $M=\SI{40}{kg}$ and $k=\SI{8000}{N.m^{-1}}$, $N_0=\SI{196.2}{N}$ and $\delta_0=\SI{24.525}{mm}$. The equilibrium COM height is chosen as \SI{1.5}{m}; the corresponding nominal support height is $b_0=z_0-h+\delta_0$. Case B adds a 0.5-degree initial roll perturbation to this equilibrium height to test settling; the other representatives start at zero roll. All velocities, controller states and work/dissipation accumulators start at zero. Steady metrics exclude the initial four seconds.

\section{Small-angle limit, stability and transfer functions}

Write $z=z_0+\eta$, $b_i=b_0+\beta_i$, and expand around $\theta=0$. To first order,
\begin{align}
z_i&=z_0-h+\eta+s_i\theta+O(\theta^2),\\
J_i&=s_i+h\theta+O(\theta^2),\\
N_i&=N_0+k(\beta_i-\eta-s_i\theta)
 +c(\dot\beta_i-\dot\eta-s_i\dot\theta).
\end{align}
Substitution in equation~\eqref{eq:finite}, using $s_L+s_R=0$ and $s_L^2+s_R^2=2a^2$, yields
\begin{align}
M\ddot\eta+2c\dot\eta+2k\eta
 &=k(\beta_R+\beta_L)+c(\dot\beta_R+\dot\beta_L),\label{eq:heave}\\
I\ddot\theta+2ca^2\dot\theta+(2ka^2-Mgh)\theta
 &=a\{k(\beta_R-\beta_L)+c(\dot\beta_R-\dot\beta_L)\}.
\label{eq:roll}
\end{align}
The term $Mgh$ is destabilizing because the COM lies above the support line. In this restricted passive limit, stability requires
\begin{equation}
k>0,\quad c>0,\quad K_\theta\equiv2ka^2-Mgh>0.
\end{equation}
The selected $K_\theta=\SI{3941.14}{N.m.rad^{-1}}$. Undamped natural frequencies are
\begin{equation}
f_z=\frac{1}{2\pi}\sqrt{\frac{2k}{M}}=\SI{3.1831}{Hz},\qquad
f_\theta=\frac{1}{2\pi}\sqrt{\frac{K_\theta}{I}}=\SI{3.5325}{Hz}.
\end{equation}

With common and differential inputs $\beta_+=(\beta_R+\beta_L)/2$ and $\beta_-=(\beta_R-\beta_L)/2$, the frequency-domain transfer functions are
\begin{align}
\frac{\eta(s)}{\beta_+(s)}&=\frac{2(cs+k)}{Ms^2+2cs+2k},\\
\frac{\theta(s)}{\beta_-(s)}&=\frac{2a(cs+k)}{Is^2+2ca^2s+K_\theta}.
\end{align}
These expressions apply in the small-amplitude, closed-contact limit. Tests compare steady amplitudes at reduced input; the saved-output check fits the harmonic response of B and C.

With $\beta_L=A\sin\omega t$ and $\beta_R=A\sin(\omega t+\Delta)$, sum-to-product identities give
\begin{align}
\beta_+&=A\cos(\Delta/2)\sin(\omega t+\Delta/2),\\
\beta_-&=A\sin(\Delta/2)\cos(\omega t+\Delta/2).
\end{align}
For fixed frequency, the heave transfer function is unchanged, so the matched 60-degree to zero-degree amplitude ratio is $\cos30^\circ=0.8660254038$. Recalculation from saved NPZ records gives a heave RMS ratio 0.8662219462 and a fitted harmonic amplitude ratio 0.8660251060. The identities themselves agree with the stored support deviations to $1.15\times10^{-16}$~m or better. The RMS ratio differs from the ideal harmonic ratio by 0.0227\%; the declared window includes both endpoints and subtracts its sample mean, so phase-shifted sampled sinusoids need not have precisely the same RMS-to-amplitude factor. The finite-angle responses also need not be exact sinusoids.

The harmonic fit uses columns $\sin\omega t$, $\cos\omega t$ and a constant over $4\le t\le12$~s; amplitude is the Euclidean norm of the first two coefficients. The fitted B and C amplitudes are 5.012957 and 4.341347~mm. The harmonic check reads the representative stored outputs.

For a separate force-phase diagnostic, take equal alternating force amplitude $B$:
\begin{align}
F_L&=F_0+B\sin\Omega t,\\
F_R&=F_0+B\sin(\Omega t+\Delta).
\end{align}
The alternating difference has amplitude $2B|\sin(\Delta/2)|$, so the roll-moment amplitude is
\begin{equation}
|\tau|_{\rm amp}=2aB|\sin(\Delta/2)|.
\end{equation}
The normalized equal-weight sum amplitude is $r=|\cos(\Delta/2)|$, giving $|\tau|_{\rm amp}=2aB\sqrt{1-r^2}$. A separate table evaluates this identity at 13 phase values from 0 to 180 degrees in 15-degree steps. These are analytic evaluations, not integrations, and are excluded from the 94-run count. The diagnostic concerns force phases; the simulation prescribes support displacement phases.

\subsection{Structural sensitivity and its interpretation}

Define the dimensionless gravitational-to-elastic ratio
\begin{equation}
\rho=\frac{Mgh}{2ka^2},\qquad K_\theta=2ka^2(1-\rho),\qquad
f_\theta=\frac{1}{2\pi}\sqrt{\frac{2ka^2(1-\rho)}{I}}.
\end{equation}
For positive stiffness, inertia and damping, $\rho<1$ is the passive roll-stability condition. It does not determine forced amplitude alone. The roll damping ratio and drive-to-natural-frequency ratio are
\begin{equation}
\zeta_\theta=\frac{ca^2}{\sqrt{IK_\theta}},\qquad r_\theta=\frac{f}{f_\theta}.
\end{equation}
They expose why equally stable systems can respond differently to a fixed drive.

Eight declared configurations change one quantity at a time: the reference; $a=0.20$, 0.35 or 0.65~m; $k=4000$~N/m; $I=16$~kg~m$^2$; $M=80$~kg; or $h=0.45$~m. Each uses equal 4-mm inputs at 1.5~Hz with 0- and 60-degree offsets, no feedback, a two-second ramp, 12-s duration, 1-ms integration and 10-ms output. Every run starts at its own static equilibrium with zero angle and velocity. Unlike reference case B's settling test, no initial roll perturbation is applied. Sixteen integrations are separate from the 94-run reference design. The half-spacing comparison is not inferred from the contextual mesh.

\begin{table}[!htbp]
\centering\small
\caption{Structural sensitivity over 4-12~s. $H_0,H_{60}$ are heave RMS in mm; $R_{60}$ is roll RMS in degrees. $P_{60}$ is maximum force divided by the static force per support. Only the named parameter changes. All equal-phase cases have zero roll from the symmetric initial state; all contacts remain positive throughout 0-12~s.}
\label{tab:structural}
\begin{tabular}{@{}lrrrrrr@{}}
\toprule Configuration & $\rho$ & $f_\theta$ (Hz) & $H_0$ & $H_{60}$ & $R_{60}$ & $P_{60}$\\\midrule
Reference & 0.0147 & 3.533 & 3.543 & 3.069 & 0.1966 & 1.044\\
$a=0.20$ m & 0.0920 & 1.356 & 3.543 & 3.069 & 1.1825 & 1.329\\
$a=0.35$ m & 0.0300 & 2.453 & 3.543 & 3.069 & 0.3577 & 1.066\\
$a=0.65$ m & 0.0087 & 4.606 & 3.543 & 3.069 & 0.1392 & 1.041\\
$k=4000$ N/m & 0.0294 & 2.479 & 4.051 & 3.508 & 0.2209 & 1.051\\
$I=16$ kg m$^2$ & 0.0147 & 2.498 & 3.543 & 3.069 & 0.2430 & 1.062\\
$M=80$ kg & 0.0294 & 3.506 & 4.684 & 4.057 & 0.1998 & 1.054\\
$h=0.45$ m & 0.0441 & 3.479 & 3.543 & 3.069 & 0.2031 & 1.046\\
\bottomrule
\end{tabular}
\end{table}

At $a=0.20$~m, $K_\theta=581.14$~N~m/rad, $f_\theta=1.3565$~Hz and $\zeta_\theta=0.1467$. Relative to the reference, the narrower spacing reduces rotational stiffness and damping and places the 1.5-Hz drive closer to the natural frequency. Roll increases about sixfold despite $\rho$ remaining well below one. Doubling inertia leaves $\rho$ unchanged but alters frequency and amplitude, so it cannot serve as a single robustness score. The smallest force over all 16 full trajectories is 131.812~N; maximum work-energy residual is $2.65\times10^{-9}$~J or less. The closed-contact assumptions used for the modal comparison hold in these numerical cases.

The structural finding is the common/differential separation in the symmetric linear limit. Response magnitude, natural frequency and contact load are parameter-dependent findings. Step-size convergence is a distinct measure of numerical accuracy for selected parameters. None is an estimate of between-person variability or a calibrated uncertainty interval for actual apparatuses. 

\section{Energy and piecewise dissipation}

Define
\begin{equation}
E=\frac12M\dot z^2+\frac12I\dot\theta^2+Mgz
+\sum_i\frac12k_i[\delta_i]_+^2,
\qquad [x]_+=\max(x,0).
\end{equation}
Away from switching instants, multiplying the equations of motion by generalized velocities shows that body power is $\sum_iN_i\dot z_i$. On a compressed branch, elastic power is $k_i\delta_i\dot\delta_i=k_i\delta_i(\dot b_i-\dot z_i)$. Therefore
\begin{align}
\dot E&=\sum_i\left[N_i\dot z_i+k_i\delta_i(\dot b_i-\dot z_i)\right],\\
\sum_iN_i\dot b_i-\dot E
&=\sum_i(N_i-k_i\delta_i)\dot\delta_i\equiv\dot D\geq0.\label{eq:energy}
\end{align}
If the raw force is positive, $N_i-k_i\delta_i=c_i\dot\delta_i$ and the term is $c_i\dot\delta_i^2$. If compression remains positive but the raw force is clipped to zero, the term is $-k_i\delta_i\dot\delta_i$, which is non-negative because clipping there requires sufficiently rapid unloading. For a gap, both force and stored contact energy are zero. The code evaluates this piecewise expression and clips only floating-point roundoff below zero.

The accumulated audit is
\begin{equation}
R_E(t)=E(t)-E(0)-W_s(t)+D(t),\qquad
W_s=\int_0^t\sum_iN_i\dot b_i\,dt.
\end{equation}
Positive and negative parts of each $N_i\dot b_i$ are integrated separately. In the response map the maximum absolute residual was approximately \SI{1.3e-6}{J}; in representative \SI{1}{ms} runs it was about \SI{1e-9}{J} or smaller.

\section{Support inputs and the bounded feedback comparison}

Every open-loop support height has the form
\begin{equation}
b_i(t)=b_0+r(t)\,x_i(t),
\end{equation}
where $r$ is a two-second raised-cosine ramp, distinct from the structural ratio $\rho$. Periodic cases use $x_i=A_i\sin(2\pi ft+\phi_i)$. The irregular case uses fixed sums of 0.45, 0.9 and \SI{1.7}{Hz} components with separately declared phases and amplitudes. It is bandwidth-limited deterministic input; the three additional phase sets assess synthetic waveform dependence and are not samples of people.

For feedback case D, $b_i=b_0+r x_i+u_i$. A state $q_i$ filters local deviation from static load,
\begin{equation}
\dot q_i=\frac{N_i-N_0-q_i}{T_q},\qquad T_q=\SI{0.08}{s}.
\end{equation}
The controller is
\begin{equation}
\dot u_i=\operatorname{clip}(-g_qq_i-u_i/T_u,-v_{\max},v_{\max}),
\end{equation}
subject to the displacement bound $|u_i|\leq u_{\max}$. Values are $g_q=\SI{5e-5}{m.N^{-1}.s^{-1}}$, $T_u=\SI{1}{s}$, $v_{\max}=\SI{0.02}{m.s^{-1}}$ and $u_{\max}=\SI{0.006}{m}$. The current command depends on the filtered state, while force is resolved from the evolving mechanical state, avoiding an instantaneous command-force algebraic loop. Yielding under above-static force is the declared sign. This is a synthetic controller, not a measured human response. The term yielding does not establish controller passivity: the contact dissipation audit in S7 is not a passivity proof for the commanded support. No anticipation, visual pathway or muscle co-contraction is represented. The small effect of case D tests this rule and parameter choice only.

\section{Reference design and complete parameter table}

\begin{longtable}{@{}p{0.39\linewidth}p{0.18\linewidth}p{0.16\linewidth}p{0.18\linewidth}@{}}
\caption{Values in \code{experiments.yaml}. All numerical values are assumed unless labeled calculated.}\label{tab:fullparams}\\
\toprule
Quantity & Value & Units & Status\\
\midrule
\endfirsthead
\toprule
Quantity & Value & Units & Status\\
\midrule
\endhead
Mass $M$ & 40 & kg & assumed\\
Roll inertia $I$ & 8 & kg m$^2$ & assumed\\
Half-spacing $a$ & 0.5 & m & assumed\\
COM height $h$ & 0.15 & m & assumed\\
Stiffness $k$ per support & 8000 & N m$^{-1}$ & assumed\\
Damping $c$ per support & 250 & N s m$^{-1}$ & assumed\\
Gravity $g$ & 9.81 & m s$^{-2}$ & fixed convention\\
Equilibrium COM height & 1.5 & m & assumed\\
Static force per support & 196.2 & N & calculated\\
Static compression & 0.024525 & m & calculated\\
Input ramp & 2 & s & assumed\\
Run/analysis interval & 12 / 4-12 & s & design\\
Integrator/log step & 0.001 / 0.01 & s & design\\
Periodic amplitude/frequency & 0.004 / 1.5 & m / Hz & assumed\\
Case C,D phase gap & 60 & degrees & assumed\\
Case B initial roll & 0.5 & degrees & test perturbation\\
Map frequencies & 0.5-5 (9 points) & Hz & design\\
Map phase gaps & 0-180 (7 points) & degrees & design\\
Asymmetry ratios & 0.6-1.4 (5 points) & - & design\\
Damping multipliers & 0.5, 1, 2 & - & design\\
Feedback filter/return & 0.08 / 1.0 & s & assumed\\
Feedback gain & $5\times10^{-5}$ & m N$^{-1}$ s$^{-1}$ & assumed\\
Feedback displacement/rate limits & 0.006 / 0.02 & m / m s$^{-1}$ & assumed\\
\bottomrule
\end{longtable}

The frequency grid is $(0.5,1,1.5,2,2.5,3,3.5,4,5)$~Hz and the phase grid is $(0,30,60,90,120,150,180)$ degrees. Sensitivity uses right/left amplitude ratios $(0.6,0.8,1,1.2,1.4)$, in-phase inputs and damping multipliers $(0.5,1,2)$. The irregular input uses frequency vector $(0.45,0.9,1.7)$~Hz and left/right amplitude vectors $(0.8,0.6,0.4)$ and $(0.7,0.65,0.45)$~mm. Its representative phase vectors are $(0.2,2.0,4.1)$ and $(1.1,3.2,5.0)$ radians. The three extra phase-set runs use that pair and the pairs $(0.7,2.7,5.3)/(1.8,4.0,0.4)$ and $(1.2,3.8,0.8)/(2.4,4.7,1.5)$, without stochastic draws. The two-second ramp is $r(t)=[1-\cos(\pi t/2)]/2$ for $0<t<2$, zero before and one thereafter.

All representative RMS statistics use the 801 logged samples with $4\le t\le12$~s at 0.01-s spacing. Heave is demeaned over that window, roll is measured about zero, and force variation is measured about $N_0$ per side. The reported aggregate force-variation comparison averages the two per-side RMS values. Contact loss is the fraction of samples with $N_i\le10^{-9}$~N. The response-map, representative, sensitivity and phase-set runs use a 1-ms step, and the convergence runs use their explicitly varied step. The log interval remains 0.01~s. No stochastic population sampling or statistical confidence intervals are implied.

The reference run count is 94: four representatives, 63 response-map cells, 15 asymmetry/damping cells, three irregular phase sets and nine convergence runs (three cases at three time steps). The time-step study reuses identical analytic input functions and controller time constants. It does not compare different random histories.

\section{Complete numerical results and sensitivities}

\begin{table}[ht]
\centering
\small
\caption{Four representative comparisons. RMS values use the declared post-transient window.}
\label{tab:cases}
\begin{tabular}{@{}lrrrr@{}}
\toprule
Case & Heave RMS (mm) & Roll RMS (deg) & Peak force / static & Contact loss\\
\midrule
A: irregular & 0.6981 & 0.04442 & 1.0094 & 0\\
B: common 1.5 Hz & 3.5425 & $<10^{-12}$ & 1.0454 & 0\\
C: 60-degree offset & 3.0686 & 0.19659 & 1.0438 & 0\\
D: C plus load yielding & 3.0520 & 0.19572 & 1.0436 & 0\\
\bottomrule
\end{tabular}
\end{table}

Relative to case C, case D changed roll RMS by $-0.4457\%$ and the RMS variation of support force by $-0.5243\%$. This weak result is preserved because the selected controller was causal yet did little under the declared parameters. It neither identifies human feedback nor tests all possible load-responsive policies.

\begin{figure}[!htbp]
\centering
\includegraphics[width=\linewidth]{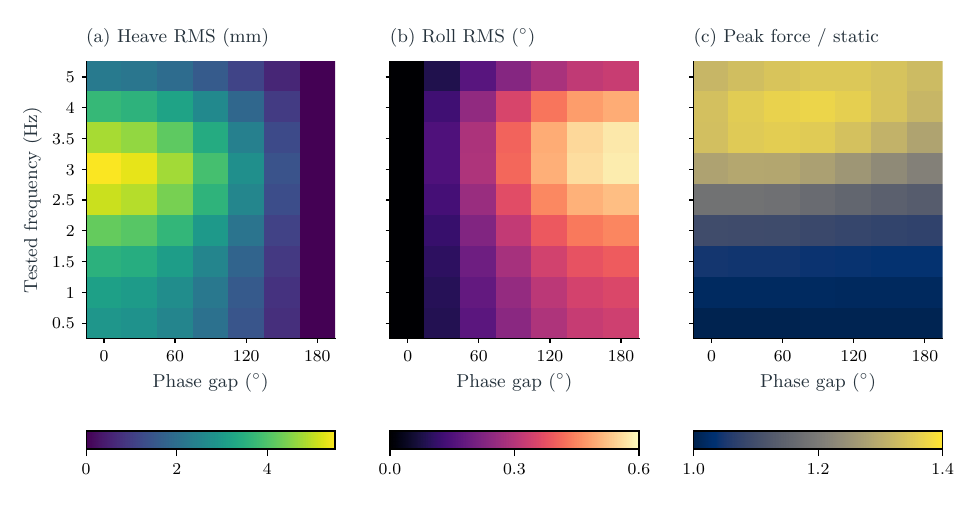}
\caption{Complete 63-cell synthetic response map. The three panels use separate colour scales and physical units. Inputs are equal-amplitude 4-mm sinusoids; frequencies and phase gaps are prescribed. Each tile denotes one tested pair. Rows are labelled by their tested frequency; they are not a uniformly spaced continuous frequency axis. No interpolation is implied.}
\label{fig:map}
\end{figure}

In the 63-cell map, maximum heave was \SI{5.4756}{mm} RMS at \SI{3}{Hz} and zero phase gap. Maximum roll was \SI{0.57868}{degree} RMS at \SI{3}{Hz} and 180 degrees; it occurred in a different cell. The maximum peak support force was 1.3707 times static load at \SI{4}{Hz} and 90 degrees. No cell lost contact. Figure~\ref{fig:map} shows the complete map with separate numerical scales for dimensionally different quantities.

\begin{figure}[!htbp]
\centering
\includegraphics[width=\linewidth]{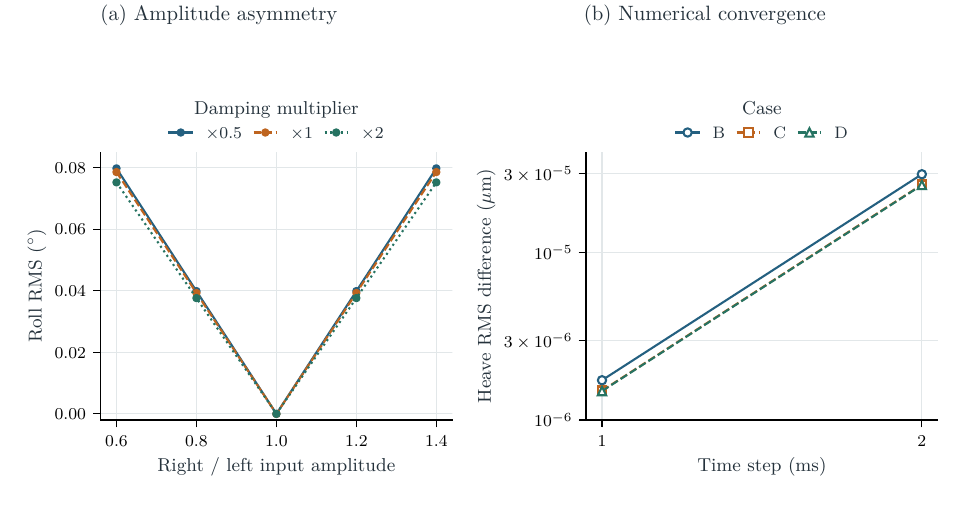}
\caption{Sensitivity and numerical checks. Left: roll under input-amplitude asymmetry for three damping multipliers. Right: absolute heave-RMS difference relative to the 0.5-ms step. Only the 1- and 2-ms comparisons are plotted: the zero difference of the reference itself is omitted from the logarithmic axis, rather than replaced by a positive floor. B, C and D denote the representative cases in Table~\ref{tab:cases}. These synthetic checks do not measure between-carrier variation.}
\label{fig:checks}
\end{figure}

The asymmetry calculation changes the right input amplitude relative to the left at three damping levels. It confirms that even nominally common timing can create roll when amplitudes differ. The time-step comparison covers cases B-D at 2, 1 and \SI{0.5}{ms}; reported RMS metrics are stable beyond the significant figures used in the main text. Figure~\ref{fig:checks} separates this sensitivity from the numerical check.

\FloatBarrier

\section{Numerical implementation and checks}

The state contains $(z,\theta,\dot z,\dot\theta)$, two feedback displacements, two filtered loads, positive work, negative work and dissipation. A fixed-step classical fourth-order Runge-Kutta method advances the ordinary differential equation. The gated force is evaluated at every RK stage. The representative log interval is ten integration steps.

Eight focused tests verify the reference solver. They cover symmetric equilibrium/preload, torque sign from differential forcing, absence of force across a gap, torque cancellation at symmetric input, the force-phase identity, small-signal amplitude agreement, moving-support work balance and finite energy. Convergence runs separately compare 2, 1 and \SI{0.5}{ms} steps. These checks are not evidence about carrier populations.
For example, case C roll RMS is 0.003431222930, 0.003431222899 and \SI{0.003431222897}{rad} over those three steps. The residual decreases at roughly fourth order while output metrics are already stable at the reported precision. All focused tests pass in the project environment.

\section{Replay verification and additional apparatus views}

The reference finite-angle RK4 solver saves position and quaternion in $(w,x,y,z)$ order. Blender reads these arrays without a rigid-body world or any additional dynamics. Both numerical and rendering axes are $x$ fore-aft, $y$ lateral and $z$ up. The illustrative rath retains its original 39 components and follows the saved rigid pose; its graphical origin does not identify a measured COM. A separate analytical replay displays the numerical contacts and forces. The videos illustrate the two-support reference cases. The walking extension saves poses in the same world-axis and quaternion conventions; its interactive browser display is a separate implementation, not a Blender replay or an additional dynamics engine.

Evaluated Blender world transforms are read back and compared with the numerical position and normalised quaternion. Position error is Euclidean distance; orientation error is $2\cos^{-1}(\min\{1,|\mathbf q^T\widetilde{\mathbf q}|\})$, accounting for quaternion sign equivalence. The illustrative 30-s replay passed on 126 distinct frames, including 120 seeded random frames and endpoints/extrema: maximum position error $5.70\times10^{-8}$~m and orientation error $2.98\times10^{-8}$~rad, below $2\times10^{-6}$ tolerances. Component-transform and geometry-hash checks also passed. These are transfer checks, not calibration against a procession.

Each condition shows saved 4-12~s over 15~s of screen time, with identical mapping, camera and unamplified displacement. The replay's numerical metre coordinate does not give the illustrated object an empirical scale. The main numerical results do not depend on viewing the animation.

\begin{figure}[!htbp]
\centering
\includegraphics[width=\linewidth]{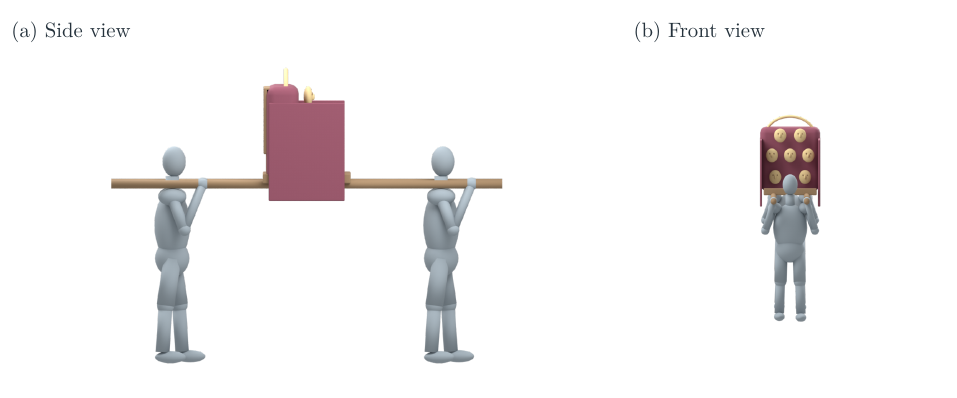}
\caption{Additional (a) side and (b) front views of the original contextual apparatus in main Figure~1. The front and rear silhouettes overlap in the front view. Proportions, hidden contacts and posture are illustrative; these views do not supply measured dimensions.}
\label{fig:additionalviews}
\end{figure}

\section{Walking-informed inputs and four-contact mechanics}

\subsection{Evidence, assumptions and phase convention}

The extension represents walking through prescribed shoulder heights rather than articulated legs or foot-ground forces. The periodicity choice follows the distinction between twice-per-stride vertical motion and once-per-stride frontal-plane motion reported by \citet{thorstensson1984}; the publisher abstract was inspected, not a digitised shoulder recording. The cadence contrasts use values appearing in the pole-carrying study of \citet{castillo2014}; relevant publisher-indexed methods and results were inspected. Neither study calibrates this apparatus. The complete publisher text of \citet{lanini2017}, especially Sections~2.1-2.2, motivates keeping prescribed timing distinct from adaptation during paired hand-carrying. No dataset from those studies was reanalysed. The separate source and parameter ledger records these access limits.

For front and rear carriers $j\in\{F,R\}$, let
\begin{align}
\alpha_j(t)&=\pi f_{\mathrm{step},j}t+\chi_j
 +m_j\sin(2\pi f_m t+\nu_j),\\
v_j(t)&=A_j\cos(2\alpha_j),\qquad
\psi_j(t)=\Psi_j\sin\alpha_j,\\
b_{j,s}(t)&=b_0+r(t)\{v_j(t)+s\sin\psi_j(t)\},\qquad s\in\{-a,+a\}.
\end{align}
Here $r$ is the same two-second raised-cosine ramp as in S8; $v_j$ is the midpoint-height deviation and $\psi_j$ the prescribed torso tilt. A stride is a complete two-step cycle, so $\dot\alpha_j/(2\pi)$ is stride frequency and twice that value is instantaneous step frequency. The phase origin and within-stride timing of vertical motion relative to tilt are declared waveform assumptions. This is not a model of footfall impact or stance duration. The analytic derivatives of $\alpha_j$, $v_j$, $\psi_j$ and $r$ give $\dot b_{j,s}$; no differentiated noisy samples are used.

The front-rear stride offset $\Gamma=\chi_R-\chi_F$ differs from the reference pole-side phase $\Delta$. At $\Gamma=\pi/2$, the vertical midpoint inputs are opposed; at $\Gamma=\pi$, they coincide while torso tilts oppose. Torso tilt represents a shoulder-height difference, not calculated lateral translation or a horizontal force. The model does not choose, entrain or adapt its own stepping rhythm. A changed cadence or phase is an imposed counterfactual input.

\subsection{Coordinates and finite-angle dynamics}

\begin{figure}[!htbp]
\centering\includegraphics[width=\linewidth]{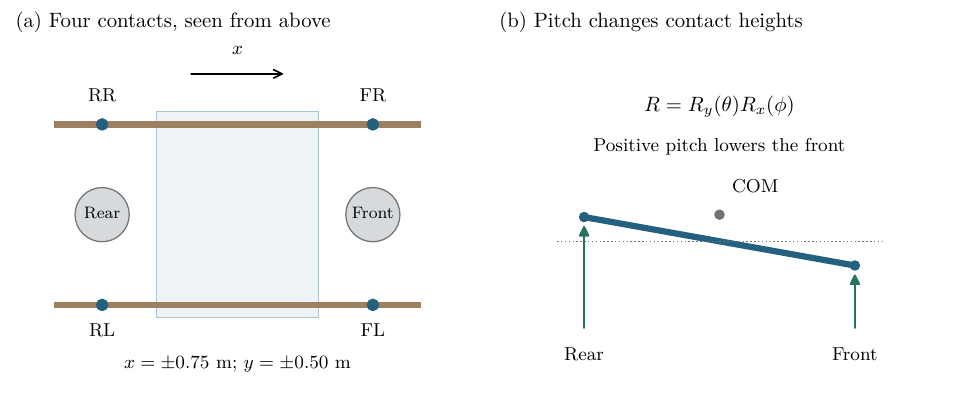}
\caption{Geometry of the walking extension. (a) Front and rear pairs of left/right contacts, with assumed coordinates. (b) Positive pitch lowers front body contacts under the declared rotation convention. The pitch is exaggerated to explain the sign; dimensions and inertia are assumptions, not measurements from the contextual apparatus. Only vertical normal forces are retained.}
\label{fig:walkinggeometry}
\end{figure}

To distinguish the two angles, this extension denotes roll by $\phi$ and pitch by $\theta$; the reference roll $\theta$ in S3-S11 is thus renamed $\phi$ here. Generalized coordinates are $(z,\phi,\theta)$ and orientation is $R_y(\theta)R_x(\phi)$. The centre of mass remains at $(0,0,z)$; yaw is fixed in this rotation convention. Body-frame contact points are $(x_i,y_i,-h)$, with $x_i=\pm\ell$, $y_i=\pm a$. Columns are front-left, front-right, rear-left, rear-right; left is negative $y$ (Figure~\ref{fig:walkinggeometry}). Define
\begin{align}
u_i&=y_i\sin\phi-h\cos\phi,\\
z_i&=z-x_i\sin\theta+u_i\cos\theta,\\
A_i^{q}&=\partial z_i/\partial\phi
=\cos\theta(y_i\cos\phi+h\sin\phi),\\
B_i^{q}&=\partial z_i/\partial\theta
=-x_i\cos\theta-u_i\sin\theta.
\end{align}
The superscript distinguishes geometric Jacobians from the input amplitudes $A_j$. Contact velocity is $\dot z_i=\dot z+A_i^q\dot\phi+B_i^q\dot\theta$. With $\delta_i=b_i-z_i$, the force law remains
\begin{equation}
N_i=\begin{cases}\max(0,k_p\delta_i+c_p\dot\delta_i),&\delta_i>0,\\0,&\delta_i\leq0.\end{cases}
\end{equation}
The body angular velocity is $(\dot\phi,\dot\theta\cos\phi,-\dot\theta\sin\phi)$ in body coordinates. Thus, with $J(\phi)=I_y\cos^2\phi+I_z\sin^2\phi$, the kinetic energy is
\begin{equation}
T=\tfrac12M\dot z^2+\tfrac12I_x\dot\phi^2+\tfrac12J(\phi)\dot\theta^2.
\end{equation}
Lagrange's equations give
\begin{align}
M\ddot z&=\sum_iN_i-Mg,\\
I_x\ddot\phi&=\sum_i A_i^qN_i+\tfrac12J'(\phi)\dot\theta^2,\\
J(\phi)\ddot\theta&=\sum_i B_i^qN_i-J'(\phi)\dot\phi\dot\theta,
\end{align}
where $J'=2(I_z-I_y)\sin\phi\cos\phi$. Including these inertia terms avoids treating finite roll and pitch as two independent planar rotations. Horizontal and yaw constraints can support reaction loads; they perform no work in the retained generalized coordinates. No horizontal contact force, flexural degree of freedom or active human controller is implied.

The energy is $E=T+Mgz+\tfrac12k_p\sum_i[\delta_i]_+^2$. The S7 work identity applies patch by patch: boundary power is $\sum_iN_i\dot b_i$ and dissipation is $\sum_{\delta_i>0}(N_i-k_p\delta_i)\dot\delta_i\geq0$. The solver integrates net boundary work and dissipation with the mechanical state. At the symmetric equilibrium, the linear angular stiffnesses are $4k_pa^2-Mgh$ and $4k_p\ell^2-Mgh$. With identical front/rear drives and zero initial pitch, the roll/heave equations and summed side forces recover the original two-support system exactly. Saved reference cases B and C provide a numerical check of that reduction.

\subsection{Parameters and comparison modes}

Mass, roll inertia, lateral half-spacing, COM offset and equilibrium height retain the reference values. New assumptions are $\ell=0.75$~m, $I_y=16$ and $I_z=20$~kg~m$^2$. These principal inertias satisfy triangle inequalities but are not identified for a rath. Four patches each use $k_p=4000$~N/m and $c_p=125$~N~s/m, so the two patches on a side recover the reference coefficients. Each static force is 98.1~N; compression remains 24.525~mm and $b_0=1.374525$~m. Input amplitude is not measured tissue compression.

G0, G90 and G180 use $f_{\rm step}=2$~Hz, $A=20$~mm and $\Psi=1.5$ degrees for both carriers, with the indicated stride offset. Gvar starts at 90 degrees but sets front/rear step rates to 2.00/2.04~Hz, amplitudes to 20/16~mm and tilt amplitudes to 1.5/1.2 degrees. Its phase modulation has $m_F=m_R=0.08$~rad, $f_m=0.25$~Hz and $\nu_F=0$, $\nu_R=\pi/3$. It is one deterministic waveform, not a random sample of people. Gstress uses a 180-degree stride offset, 2.33~Hz, 35~mm and 3 degrees for both carriers. Its combined changes intentionally probe a larger-input boundary and do not isolate one causal parameter. G90slow and G90fast instead change cadence alone from G90 to 1.83 and 2.33~Hz. All other phase modulations are zero. Waveform amplitudes, relative timing, geometry and contact coefficients remain illustrative assumptions.

The full model solves all three coordinates. The pitch-locked comparison fixes $\theta=\dot\theta=0$ but retains four independent contacts. Its ideal pitch reaction performs no work. The collapsed comparison also fixes pitch and uses $\bar b_s=(b_{F,s}+b_{R,s})/2$ on each side before gating the force. Two identical half-stiffness contacts then implement each reference side resultant. Their assigned equal patch forces are bookkeeping, not predictions of individual load sharing. Comparing side sums is valid across all three models; identifying individual peaks requires separate contacts. This distinction prevents a two-to-four-contact normalization error.

\section{Walking comparisons and interpretation}

All 21 primary comparisons use RK4 with a 1-ms step for 24~s, storing poses at 0.01~s and forces at every integration step. Summaries use $8\leq t\leq24$~s inclusive: 1601 pose samples and 16001 force samples. Heave is demeaned; angular RMS values use zero as reference. Zero-force fraction pools the four contact columns and includes both geometric gaps and clipped unloading. It is a discrete contact-time fraction, not a fraction of participants or an exact event-duration estimate. For the collapsed model it describes effective-resultant unloading, not actual shoulder loss. All primary comparisons start at the static state and use the same ramp.

\begin{longtable}{@{}llrrrrr@{}}
\caption{Complete walking comparison at the primary 1-ms step. $H$, $R$, $P$: heave (mm), roll and pitch (degrees) RMS. $S$: peak side force divided by 196.2 N; $L$: zero-force contact--time percentage. Full and locked retain four patches; reduced uses two effective side resultants. Angular zeros denote values below $10^{-10}$ degrees.}\label{tab:allwalking}\\
\toprule Case & Mode & $H$ & $R$ & $P$ & $S$ & $L$ (\%)\\\midrule\endfirsthead
\multicolumn{7}{@{}l}{Table~\ref{tab:allwalking} (continued).}\\[3pt]
\toprule Case & Mode & $H$ & $R$ & $P$ & $S$ & $L$ (\%)\\\midrule\endhead
G0 & Full & 21.099 & 1.169 & 0 & 1.540 & 0.00\\*
G0 & Locked & 21.099 & 1.169 & 0 & 1.540 & 0.00\\*
G0 & Reduced & 21.099 & 1.169 & 0 & 1.540 & 0.00\\
\addlinespace[3pt]
G90 & Full & 0.046 & 0.825 & 1.427 & 1.042 & 0.00\\*
G90 & Locked & 0.342 & 0.800 & 0 & 1.087 & 11.65\\*
G90 & Reduced & 0.016 & 0.824 & 0 & 1.039 & 0.00\\
\addlinespace[3pt]
G180 & Full & 21.066 & 0 & 0 & 1.479 & 0.00\\*
G180 & Locked & 21.066 & 0 & 0 & 1.479 & 0.00\\*
G180 & Reduced & 21.066 & 0 & 0 & 1.479 & 0.00\\
\addlinespace[3pt]
Gvar & Full & 15.374 & 0.386 & 0.778 & 1.469 & 0.00\\*
Gvar & Locked & 15.381 & 0.383 & 0 & 1.469 & 1.66\\*
Gvar & Reduced & 15.374 & 0.386 & 0 & 1.469 & 0.00\\
\addlinespace[3pt]
Gstress & Full & 48.678 & 0 & 0 & 3.039 & 50.40\\*
Gstress & Locked & 48.678 & 0 & 0 & 3.039 & 50.40\\*
Gstress & Reduced & 51.501 & 0 & 0 & 3.264 & 52.07\\
\addlinespace[3pt]
G90slow & Full & 0.050 & 0.810 & 1.365 & 1.035 & 0.00\\*
G90slow & Locked & 0.400 & 0.787 & 0 & 1.075 & 11.41\\*
G90slow & Reduced & 0.015 & 0.810 & 0 & 1.033 & 0.00\\
\addlinespace[3pt]
G90fast & Full & 0.041 & 0.845 & 1.552 & 1.056 & 0.00\\*
G90fast & Locked & 0.280 & 0.819 & 0 & 1.113 & 11.94\\*
G90fast & Reduced & 0.020 & 0.845 & 0 & 1.053 & 0.00\\
\bottomrule
\end{longtable}

At 90 degrees of stride separation, front/rear vertical inputs oppose, exciting pitch that the reference reduction excludes. At 180 degrees, diagonal differences in prescribed torso tilt cancel net angular loads in the symmetric configuration, although individual contact forces change. In that closed-contact case, side sums and heave agree with the reduced model while equal allocation conceals the largest individual load. The larger-input case opens contacts, and averaging then fails even to preserve side-resultant dynamics. These observations have different causes: missing pitch, missing individual load information, and nonlinear contact aggregation should not be treated as a single failure of the planar model.

The G90 cadence comparison changes pitch RMS from 1.365 to 1.552 degrees and peak patch load from 1.575 to 1.781 times static over the tested rates; all its full-model contacts remain loaded. Locking pitch gives nonzero zero-force fractions at all three rates. The original small-input reference cases remain valid examples of their stated scope. This extension supports neither a calibrated risk assessment nor a universal claim that asynchronous walking destabilizes a carried object.

\clearpage
\begingroup\small
\setlength{\bibsep}{2pt}

\endgroup